\documentclass[pre,amsmath]{revtex4}
\usepackage{graphicx}
\usepackage{color}
\usepackage{bm}
\usepackage{epsfig}
\usepackage{latexsym}
\usepackage{pifont}
\usepackage{float}
\usepackage[utf8]{inputenc}
\graphicspath{{figure/}}
\usepackage{siunitx}

\begin{document}

\newcommand{\be}{\begin{equation}}
\newcommand{\ee}{\end{equation}}
\newcommand{\bea}{\begin{eqnarray}}
\newcommand{\eea}{\end{eqnarray}}
\newcommand{\nn}{\nonumber}

\title{Dynamics of coupled modes for sliding particles on a fluctuating landscape}
\author{Shauri Chakraborty(1), Sakuntala Chatterjee(1) and Mustansir Barma(2)}
\affiliation{(1) Department of Theoretical Sciences, S. N. Bose National Centre for Basic Sciences, Block JD, Sector 3, Salt Lake, Kolkata 700106, India. \\ 
(2) TIFR Centre for Interdisciplinary Sciences, Tata Institute of Fundamental Research, Gopanpally, Hyderabad 500107, India.}
\begin{abstract}
The recently developed formalism of nonlinear fluctuating hydrodynamics (NLFH) has been instrumental in unraveling many new dynamical universality classes in coupled driven systems with multiple conserved quantities. In principle, this formalism requires knowledge of the exact expression of locally conserved current in terms of local density of the conserved components. However, for most nonequilibrium systems an exact expression is not available and it is  important to know what happens to the predictions of NLFH in these cases. We address this question for the first time here in a system with coupled time evolution of sliding particles on a fluctuating energy landscape.  In the disordered phase this system shows short-ranged correlations,  this system shows short-ranged correlations, the exact form of which is not known, and so the exact expression for current cannot be obtained. We use approximate expressions based on mean-field theory and corrections to it, to test the prediction of NLFH using numerical simulations. In this process we also discover important finite size effects and show how they affect the predictions of NLFH. We find that our system is rich enough to show a large variety of universality classes. From our analytics and simulations we have been able to find parameter values which lead to diffusive, Kardar-Parisi-Zhang (KPZ), $5/3 $ L\'evy and modified KPZ universality classes. Interestingly, the scaling function in the modified KPZ case turns out to be close to the Pr\"ahofer-Spohn function which is known to describe usual KPZ scaling. Our analytics also predict the golden mean and the $3/2$ L\'evy universality classes within our model but our simulations could not verify this, perhaps due to strong finite size effects. 
\end{abstract}
\maketitle
\section{Introduction}
\label{intro}

Recently there has been a surge of research interest in uncovering various different dynamical universality classes that show up in nonequilibrium systems with more than one conserved component \cite{das01pre2, henk12, spohn14, spohn13, spohn15jsp, mukamel, sasamoto18}. In several cases, the coupled time evolution of these conserved fields gives rise to ballistically moving modes with slow decays that govern the large scale properties of the system. Using the formalism of nonlinear fluctuating hydrodynamics (NLFH), it has been shown that the spatio-temporal fluctuations of each mode can be described by a dynamical exponent $z$ and a universal scaling function that does not depend in detail on the microscopic properties of the system \cite{spohn13,popkov16jsm}. Power laws describing slow decay of the modes are associated with different universality classes which crucially depend on the nature of coupling between the modes. More specifically, how the time evolution of a particular slow mode is affected by other slow modes at the non-linear (quadratic) level determines its universality class. NLFH shows that this coupling can give rise to new values of $z$, which are different from the commonly encountered values, $z=2$ and $3/2$ for diffusive and Kardar-Parisi-Zhang (KPZ) universality classes, respectively. So far in various different driven diffusive systems $5/3$-L\'evy, $3/2$-L\'evy and golden mean universality classes have been observed \cite{gunter14, gunter15, spohn15jsp}. A $5/3$-L\'evy universality class is characterized by $z=5/3$ and a scaling function given by L\'evy $5/3$-stable distribution. Similarly,  $3/2$-L\'evy and golden mean universality classes have L\'evy $z$-stable distributions as scaling functions with $z=3/2$ and $(\sqrt{5}+1)/2$, respectively. In a particularly interesting development in this direction, it was shown that the possible values of $z$ can be expressed as the Kepler ratio of successive numbers of the Fibonacci sequence \cite{popkov15pnas}.

In this paper, we consider a coupled system consisting of sliding particles on a fluctuating potential energy landscape in one dimension. The particles tend to slide down the local potential gradient of the landscape, towards the region of minimum potential energy. In addition they also tend to modify the local dynamics of the landscape. The dynamics of the system is such that the density of the particles and the local height gradient of the landscape are conserved. In an earlier study \cite{chakraborty16prer} we have shown that by changing the coupling parameters between the particle dynamics and landscape dynamics, one can obtain a rich phase diagram which consists of various different ordered and disordered phases. The detailed characterization of the static and dynamic properties of the different ordered phases was done in \cite{chakraborty17pre1, chakraborty17pre2}. Here, we focus on the disordered phase. Specifically, we are interested in how the coupled time evolution of particle density and landscape height gradient gives rise to different dynamical universality classes in the system, following the prescription of NLFH.

We find that our system is rich enough to show various different universality classes. Unlike most NLFH studies so far, the exact steady state measure is not known for our system. Although in the disordered phase neither the particles nor the landscape show any long-range order, there are still short-ranged correlations present in the system, the exact form of which is not known, and hence exact expressions for the locally conserved currents in terms of the conserved densities remain unavailable. Therefore, we use approximate expressions based on mean-field theory where we neglect all correlations between sites or a slightly improved approximation where we retain some nearest neighbor or next nearest neighbor correlations and ignore the rest. Using these approximate expressions for the current, we carry out the NLFH analysis. In a nutshell, our procedure consists of the following steps. (i) Current-density relation: this is found analytically within mean-field theory and also within an improved approximation scheme where we keep track of neighboring correlations in a self-consistent manner. (ii) Determination of Jacobian and Hessian matrix elements: this is carried out straightforwardly using the results of (i). The results of  \cite{spohn13} and \cite{popkov16jsm} then allow us to determine parameter values of the lattice model at which new universality classes make their appearance. (iii) Numerical test: Monte Carlo simulations are carried out to determine structure functions using parameter values determined in (ii). 
\par The argument and amplitude of the structure function are scaled to obtain a data collapse, enabling an estimation of the dynamical exponent and scaling function, both of which are compared against analytic predictions. Particular care needs to be exercised to account for strong finite size effects. The scaling solution obtained from NLFH implicitly assumes the limit of infinitely large system size and time. The method of data collapse used in our simulations may be significantly  affected by finite size effects and this may even mask the actual universality class which is expected to manifest itself in the scaling limit. In this paper we explicitly demonstrate how finite size effects affect the results. We also discuss how the criteria for observing different universality classes obtained from NLFH needs to be modified in view of finite size effects.

In our paper, NLFH has been used and tested in the absence of exact knowledge of the current-density relationship, a situation which arises in many systems with coupled dynamics of conserved quantities, and where NLFH can potentially be used. We have been able to show analytically and numerically the existence of diffusive, KPZ, $5/3$-L\'evy and modified KPZ universality classes in our system. For the modified KPZ scaling \cite{schutz17arxiv} our data suggest that the  scaling function is rather close to the Pr\"ahofer-Spohn function which describes usual KPZ scaling \cite{prahofer04jsp}. Our analytics also indicate the existence of golden mean and $3/2$-L\'evy universality classes but we show how finite size effects in our system make it difficult to observe them in simulations.

In the next section, we provide a brief overview of NLFH in one dimension. In Sec. \ref{model} we define our model and present the phase diagram. In Sec.\ref{static_kw} we present results for short-ranged static correlations in the disordered phase and explain our approximation schemes to derive the expression for the current. In Sec. \ref{dynamics_kw} we present our simulation results for structure functions. Our conclusions are presented in Sec. \ref{con}.

\section{Non-linear fluctuating hydrodynamics and mode-coupling theory}
\label{framework}
The starting point for investigating the large-scale dynamical properties of a system with $n$ conserved components is the continuity equation 
\begin{equation}
\frac{\partial \vec{\rho}(x,t)}{\partial t}+\frac{\partial \vec{J}(x,t)}{\partial x}=0
\label{eq:cont1}
\end{equation}
where $\vec{\rho}(x,t)$ and $\vec{J}(x,t)$ are $n$-dimensional vectors, the components  $\rho_\alpha (x,t)$ and $J_\alpha(x,t)$ of which denote local density of the $\alpha$-th conserved quantity and associated conserved current on a mesoscopic scale, respectively, with $\alpha=1,2,...,n$. Assumption of local equilibrium ensures that the current depends on space and time only through its dependence on local densities, and does not have any explicit space-time dependence. Using this Eq. \ref{eq:cont1} can be rewritten as,
\begin{equation}
\frac {\partial \vec{\rho}}{\partial t} + \mathbf{A} \frac{\partial\vec{\rho}}{\partial x} =0
\label{eq:cont2}
\end{equation}
where $\mathbf{A}$ denotes the Jacobian with elements $A_{\alpha \beta}=\frac{\partial J_\alpha}{\partial \rho_\beta}$. Expanding the local density $\rho_\alpha (x,t)$ around its conserved global value $\rho^0_\alpha$, we write  $\rho_\alpha(x,t)=\rho_\alpha^0+u_\alpha(x,t)$. Retaining only linear terms in the  perturbation $u_\alpha(x,t)$ assumed small, we get a set of coupled linear partial differential equations that can be solved by diagonalizing $\mathbf{A}^0$, the elements of which are functions of $\{ \rho_\alpha ^0 \} $. The normal modes $\vec{\phi} = \mathbf{R}^{-1} \vec{u}$ follow the equations 
\be 
\partial_t \phi_\alpha (x,t) +\lambda_\alpha \partial_x \phi_\alpha (x,t)=0
\ee
where $\lambda_\alpha$'s are eigenvalues of $\mathbf{A}^0$. Therefore the normal modes satisfy traveling wave solutions $ \phi_\alpha (x-\lambda_\alpha t ) $ where $\lambda_\alpha$ can be interpreted as the speed of propagation of local perturbations in the system \cite{lighthill}.

Going beyond the linear theory, one can expand the current $\vec{J}$ around the stationary density values, but retain nonlinearities upto quadratic order in $\vec{u}$. This gives rise to coupling between the modes $\phi_\alpha$ in the quadratic order. The time evolution equation for $\phi_\alpha(x,t)$ then becomes \cite{henk12, popkov16jsm}
\begin{equation}
\partial_t \phi_\alpha=-\partial_x [\lambda_\alpha \phi_\alpha+ \vec{\phi}^T {\mathbf G}^\alpha \vec{\phi}-\partial_x( {\mathbf{D}}\vec{\phi})_\alpha+({\mathbf{B}} \vec{\xi})_\alpha ]
\label{eq:burger_eigen}
\end{equation}
where phenomenological diffusion and noise terms have been added \cite{mukamel}. The Gaussian white noise has the strength $\langle \xi_\alpha(x,t)\xi_\alpha(x^\prime,t^\prime) \rangle = B_{\alpha \alpha} \delta(x-x^\prime)\delta(t-t^\prime)$ and the matrix $\mathbf{B}$ can be assumed to be diagonal without any loss of generality. The mode-coupling matrices are defined as
\begin{equation}
\mathbf{G}^\alpha=\frac{1}{2}\sum_\gamma  R^{-1}_{\alpha\gamma}\mathbf{ R^T H^\gamma R}
\label{eq:Gmatrix}
\end{equation}
where the Hessian matrix $\mathbf{H}^\gamma_{\alpha \beta} = \partial^2 J_\gamma/\partial \rho_\alpha^0 \partial \rho_\beta^0$. With a knowledge of the current-density relationship, the elements of the mode-coupling matrices can be evaluated. The element $G^\alpha_{\beta \beta}$ denotes the coupling between the $\alpha$-th and $\beta$-th mode. It is easy to see from Eq. \ref{eq:burger_eigen} that the off-diagonal terms of $\mathbf{G}^\alpha$, denoted as $G^\alpha_{\beta \gamma}$ with $\beta \neq \gamma$, do not influence the time-evolution of $\phi_\alpha$. The traveling wave solution predicted from linear theory does not remain valid any more for Eq. \ref{eq:burger_eigen}, as apart from moving through the system with speed $\lambda_\alpha $, any local perturbation in $\phi_\alpha$ would also dissipate with time, due to its coupling with other modes, and also due to diffusion. The formalism of nonlinear fluctuating hydrodynamics allows us to understand the long time decay of these local fluctuations.

A useful quantity to study how local perturbations in the system decay in the limit of large space and time, is the dynamical structure function $C_{\alpha \alpha} (x,t) = \langle \phi_\alpha (0,0) \phi_\alpha (x,t) \rangle $. Starting from Eq. \ref{eq:burger_eigen} the time-evolution of $C_{\alpha \alpha} (x,t)$  can be constructed and the following scaling ansatz can be made \cite{popkov16jsm}
\begin{equation}
C_{\alpha\alpha}(x,t) \sim t^{-1/z_\alpha}f_\alpha \left (\frac{x-\lambda_\alpha t}{t^{1/z_\alpha}} \right ).
\label{eq:stf_real}
\end{equation}
Here $f_\alpha (y)$ is a scaling function and the scaling variable $y=(x-\lambda_\alpha t)/t^{1/z_\alpha}$ indicates that at time $t$ the perturbation is peaked around the position $x(t)=x(0)-\lambda_\alpha t$ while the width of the peak scales as $t^{1/z_\alpha}$. It is assumed that the spreading is sub-ballistic, {\sl i.e.} $z_\alpha > 1$. In the case when each of the eigenvalues of the matrix $\mathbf{A}^0$ is different, the modes also propagate with different speeds, in which case the cross correlation between two modes $C_{\alpha \beta} (x,t) = \langle \phi_\alpha (0,0) \phi_\beta (x,t) \rangle $ can be neglected at large times.

Taking Fourier transform in space and defining $\tilde{C}_{\alpha \alpha} (k,t) =\frac{1}{\sqrt{ 2 \pi}} \int_{-\infty}^\infty dx \exp(-ikx) C_{\alpha\alpha}(x,t)   $, we can use Eq. \ref{eq:stf_real} to write $\tilde{C}_{\alpha \alpha} (k,t) \sim e^{-i \lambda_\alpha kt} \tilde{f}_\alpha (k t^{1/z_\alpha}) $. Subsequent Laplace transform in time changes the scaling variable to  $\zeta_\alpha=(\omega+i\lambda_\alpha k){|k|}^{-z_\alpha}$ and the dynamical structure function can be written as 
\begin{equation}
\widehat{C}_{\alpha \alpha}(k,\zeta_\alpha)=\frac{1}{\sqrt{2\pi}}|k| ^{-z_\alpha}h_\alpha(\zeta_\alpha). 
\label{eq:stf_laplace}
\end{equation}
This ansatz can be used to solve the mode-coupling equation and the scaling function comes out to be \cite{popkov16jsm}
\begin{equation}
\frac{1}{h_\alpha(\zeta_\alpha)}=\lim_{k \rightarrow 0} \left [ \zeta_\alpha+D_\alpha |k|^{2-z_\alpha}+Q_{\alpha\alpha}\zeta_\alpha^{\frac{1}{z_\alpha}-1} |k|^{3-2z_\alpha}+ \sum_{\beta \neq \alpha}Q_{\alpha \beta}(-i\lambda^{\alpha \beta}_k)^{\frac{1}{z_\beta}-1} |k|^{1+\frac{1}{z_\beta}-z_\alpha} \right ]
\label{eq:scaling_fn}
\end{equation}
with coefficient $Q_{\alpha \beta}$ proportional to $(G^\alpha_{\beta \beta})^2$ and $\lambda^{\alpha \beta}_k=(\lambda_\alpha-\lambda_\beta)sgn(k)$. In order to have a nontrivial scaling limit, we must ensure that in the limit of small $k$ the scaling function $h_\alpha(\zeta_\alpha)$ stays finite and $h_\alpha(\zeta_\alpha) \neq 1/ \zeta_\alpha$. Note that $h_\alpha(\zeta_\alpha) = 1/ \zeta_\alpha$ would mean dissipationless transport of density perturbation as predicted from the linear theory. We briefly discuss below how these two criteria determine the value of the dynamical exponent $z_\alpha$ and the form of the scaling function.

\subsection{Case I} 
When all diagonal terms of $\mathbf{G}^\alpha$ vanish, $G^\alpha_{\beta \beta}=0 ~\forall \beta$, then the last two terms on the right hand side of Eq. \ref{eq:scaling_fn} drop out. The resulting scaling function will be nontrivial only if $z_\alpha =2$, which gives $h_\alpha(\zeta_\alpha)=[\zeta_\alpha+D_\alpha ]^{-1}$. This corresponds to  
\begin{equation}
\tilde{C}_{\alpha \alpha}(k,t)=\frac{1}{\sqrt{2\pi}}e^{-i\lambda_\alpha kt-D_\alpha k^2t}
\label{eq:stf_diff}
\end{equation}
This gives a diffusive universality class for the mode $\alpha$. In the absence of self-coupling and cross coupling between the modes, any local perturbation moves around the system with speed $\lambda_\alpha$ and dissipates diffusively.

\subsection{Case II}
 In case $G^\alpha_{\alpha \alpha}=0$, but there is at least one $\beta$ for which $G^\alpha_{\beta \beta} \neq 0$, then Eq. \ref{eq:scaling_fn} becomes
\begin{equation}
\frac{1}{h_\alpha(\zeta_\alpha)}=\lim_{k \rightarrow 0} \left [ \zeta_\alpha+D_\alpha |k|^{2-z_\alpha}+\sum_{\beta \neq \alpha}Q_{\alpha \beta}(-i\lambda^{\alpha \beta}_k)^{\frac{1}{z_\beta}-1} |k|^{1+\frac{1}{z_\beta}-z_\alpha}\right ]
\label{eq:scaling_fn_non_kpz}
\end{equation}
Now, we have already assumed that $z_\alpha > 1$, and hence $1+\frac{1}{z_\beta}-z_\alpha<2-z_\alpha$ which means that in the limit of small $k$ the second term in the right hand side of Eq. \ref{eq:scaling_fn_non_kpz} vanishes faster and the scaling behavior is dominated by the slowest decaying term in the summation present in the third term i.e. $z_\alpha=1+\frac{1}{{z_\beta}^{max}}$. Note that this result is consistent with the assumption that $z_\alpha > 1$. In this case the dynamic structure function in momentum space is
\begin{equation}
\tilde{C}_{\alpha\alpha}(k,t)= \frac{1}{\sqrt{2\pi}}\exp\left [-i \lambda_\alpha k t-\sum_{\beta \in \{z_\beta = z_\beta^{max}\}}Q_{\alpha \beta}(-i\lambda^{\alpha \beta}_k)^{1/z_\beta^{max}-1} |k|^{1+1/z_\beta^{max}} t \right ]
\label{eq:stf_non_kpz}
\end{equation}
Eq. \ref{eq:stf_non_kpz} shows the long time decay of the local fluctuations when mode $\alpha$ has cross-coupling with other modes but has no self-coupling term.

\subsection{Case III} 
Finally, we consider the most general case of nonvanishing self-coupling and cross coupling, $G^\alpha_{\alpha\alpha}\neq 0$ and $G^\alpha_{\beta \beta}\neq 0$ for at least one $\beta \neq \alpha$. In this case all four terms on the right hand side of Eq. \ref{eq:scaling_fn} are present. Depending on which term dominates the small $k$ behavior, we can have either $z_\alpha =2$, or $3/2$, or $1+1/z_\beta^{max}$. Thus we can rule out the possibility of $z_\alpha > 2$ even in the presence of self-coupling. Moreover, in order to make sure that the right hand side of Eq. \ref{eq:scaling_fn} does not diverge in the limit of small $k$, we must have non-negative exponents of $k$ in the diffusive term, self-coupling term and cross-coupling term. This is possible only if $z_\alpha=\min[2,3/2,1+1/z_\beta^{max}]=3/2$. The corresponding scaling function can be of two different types. If mode $\alpha$ is not cross-coupled to any diffusive mode,  i.e. $z_\beta^{max} < 2$, then only the self-coupling term dominates the small $k$ behavior and the scaling function is given by 
\begin{equation}
h_\alpha(\zeta_\alpha)=[\zeta_\alpha+Q_{\alpha\alpha}\zeta_\alpha^{-\frac{1}{3}}]^{-1}.
\label{scaling_fn_kpz}
\end{equation}
This identifies the usual KPZ universality class \cite{prahofer04jsp}. However, if $z_\beta^{max} = 2$, then the cross-coupling term also affects the scaling function and we have 
\begin{equation}
h_\alpha(\zeta_\alpha)=[\zeta_\alpha+Q_{\alpha\alpha}\zeta_\alpha^{-\frac{1}{3}}+\sum_{\beta \in \{z_\beta =2 \}}Q_{\alpha\beta}(-i\lambda_k^{\alpha\beta})^{-\frac{1}{2}}]^{-1}
\label{scaling_fn_modkpz}
\end{equation}
which is known as modified KPZ universality class \cite{spohn15jsp}. The exact scaling function for this case is not known. Not too many systems are found where modified KPZ scaling is actually observed \cite{popkov16jsm, schutz17arxiv}, but  our system displays this elusive universality class. Interestingly, our data also suggest that the form of the scaling function in this case is not too different from the usual Pr\"ahofer-Spohn scaling function for the KPZ problem. We illustrate this in Sec. \ref{sub_modkpz}. 

\section{Model and the phase diagram}
\label{model}
\begin{figure}[ht!]
\begin{center}
\includegraphics[scale=0.8]{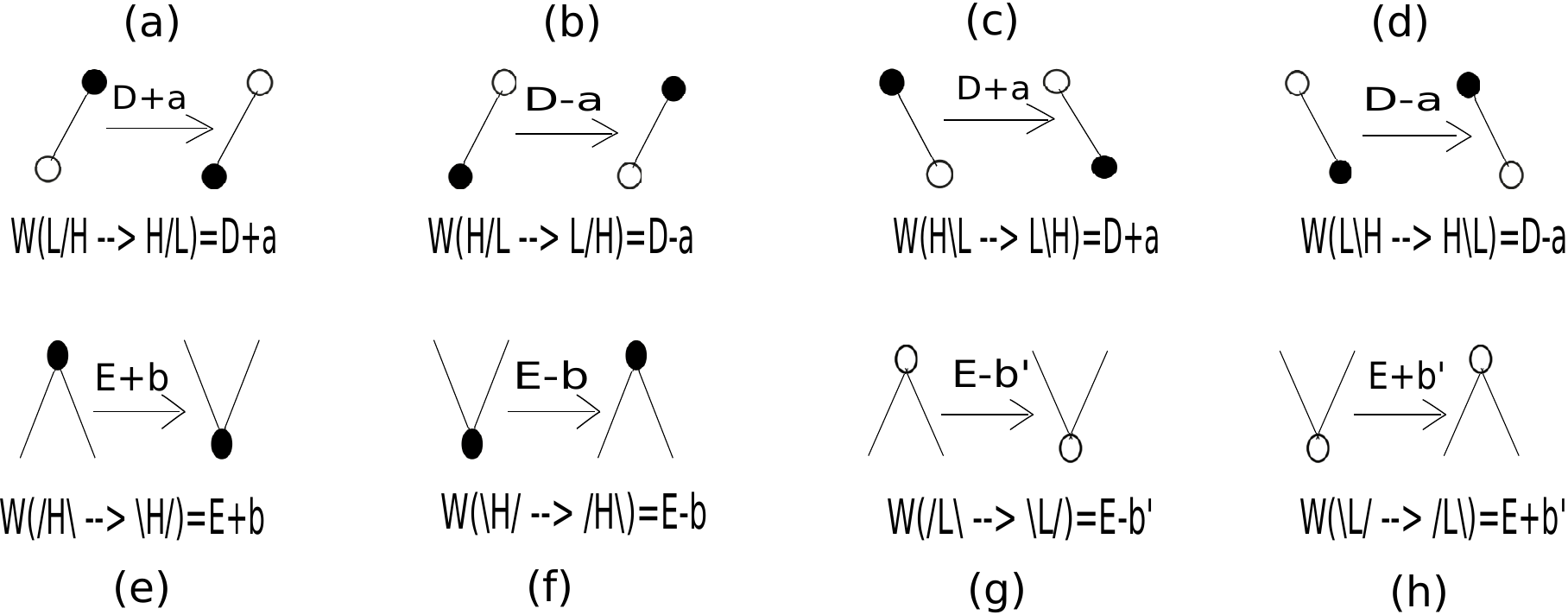}
\caption{Schematic representation of different allowed transitions in the model and the rates for each of them.  The dark (white) circles represent sites occupied by $H$ ($L$) particles. Figs. (a),(b),(c), and (d) show the rates of particle movements. An $H$($L)$-particle slides down(up) with a rate $D+a$, while the reverse moves occur with a rate $D-a$, where $D>a>0$. Figs. (e),(f),(g), and (h) show the flipping rates of local hills and valleys occupied by respectively $H$ and $L$ particles. While a local hill (valley) occupied by an $H$ can flip with a rate $E+b$ $(E-b)$, a local hill (valley) occupied by an $L$ can flip with rate $E-b^\prime$ $(E+b^\prime)$ where the parameters $b$ and $b^\prime$ can be either zero, positive, or negative  such that $E> \mid b \mid,\mid b^\prime \mid$.}
\label{model_scheme}
\end{center}
\end{figure}

Our model describes the coupled time evolution of a fluctuating landscape and particles sliding on it. The local dynamics of the landscape resembles that of a single-step model, which is  KPZ-like insofar as the time evolution happens via transition between local hills and valleys \cite{liu}. The presence of the particles affects these transition rates. The model is defined on a one dimensional lattice, each site of which can be occupied by either a heavy ($H$) particle or a light ($L$) particle. The intervening lattice bonds between two consecutive sites can have two possible orientations, an upslope bond with orientation $\pi /4$, represented using the symbol $/$ and a downslope bond with orientation $-\pi /4$, shown as $\backslash$. A combination of an upslope bond followed by a downslope bond ($/ \backslash$) is called a local hill and $\backslash /$ is a local valley. As the system undergoes time evolution, the $H$ and $L$ particles interchange their positions and   upslope and downslope bonds also switch their orientations, such that total number of $H$ (or $L$) particles and total number of upslope (or downslope) bonds are conserved in the system. We use the symbol $W(C \rightarrow C')$ to denote the transition rate from local configuration $C$ to $C'$. In Fig. \ref{model_scheme} we show these moves. We consider a total of $L$ lattice sites in our system. Out of these, a total of $N$ sites are occupied by $H$ particles and we define $\rho=N/L$. Similarly, we denote the density of upslope bonds by $m$. We use periodic boundary conditions in our system. 

\begin{figure}[ht!]
\begin{center}
\includegraphics[scale=1.5]{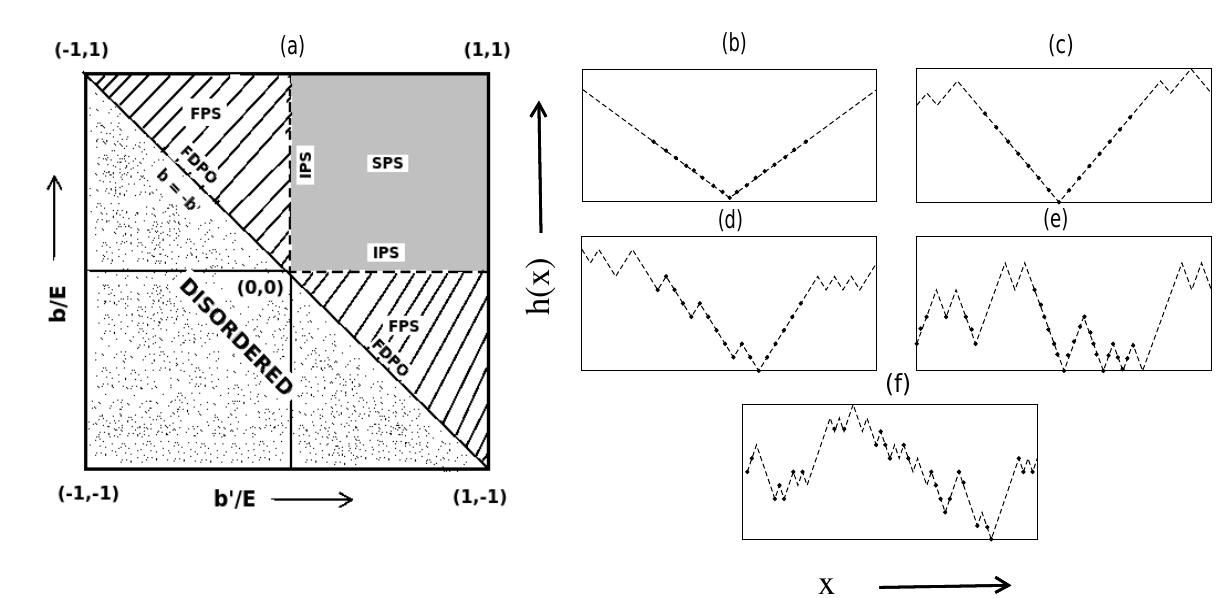}
\caption{(a) Phase diagram in the scaled $b-b'$ plane for $m=1/2$. For $b>0$ and $b'>0$, the system shows SPS (strong phase separation). On the dotted horizontal and vertical lines the system is in the IPS (infinitesimal current with phase separation) phase. The striped region ($-b<b'<0$) in the second and fourth quadrants represent the FPS (finite current with phase separation) phase. The $b=-b'$ line corresponds to FDPO (fluctuation dominated phase ordering) phase. The dotted region below the $b=-b'$ line corresponds to the disordered phase. This phase diagram is valid for all $\rho$. (b)-(f): Representative configurations for SPS, IPS, FPS, FDPO and disordered phases, respectively.}
\label{phase_d}
\end{center}
\end{figure}

In earlier studies \cite{chakraborty16prer, chakraborty17pre1, chakraborty17pre2} we had presented a phase diagram for the system by varying the rate parameters $b$ and $b'$ for a fixed value of $a$, while the phase diagram obtained on varying $a$ and $b$ when $b=b'$ was given in \cite{das01pre2}. This phase diagram contains a number of nonequilibrium ordered and disordered phases, as shown in Fig. \ref{phase_d}. Among the various kinds of ordered phases, we have strong phase separation, infinitesimal current with phase separation, and finite current with phase separation. In these three phases, the landscape shows a long range ordered phase where upslope or downslope bonds phase separate completely from each other, resulting in a large deep valley in the system.  The $H$ particles are present in the lower portion of the valley in a compact cluster which has a macroscopic extent as the particles obey a hard-core constraint. In \cite{chakraborty17pre1, chakraborty17pre2} we discussed the static and dynamic properties of the ordered phases in detail. The $b=-b'$ line acts as the boundary between the ordered and disordered phase and on this line fluctuation dominated phase ordering is observed, where landscape is completely disordered but the $H$ particles show a tendency to form large clusters of fluctuating lengths \cite{das00prl,das01pre1,chatterjee06pre,kapri16pre}. In the disordered phase, neither the particles nor the landscape show any long ranged order. In this paper, we focus on the disordered phase. We are particularly interested in applying the idea of NLFH to the coupled time evolution of two conserved densities of $H$-particles and upslope bonds to explore different dynamical universality classes present in the system. 


\section{Short-ranged correlations and mean-field calculation in the disordered phase}
\label{static_kw}
\begin{figure}[ht!]
\includegraphics[scale=1.5]{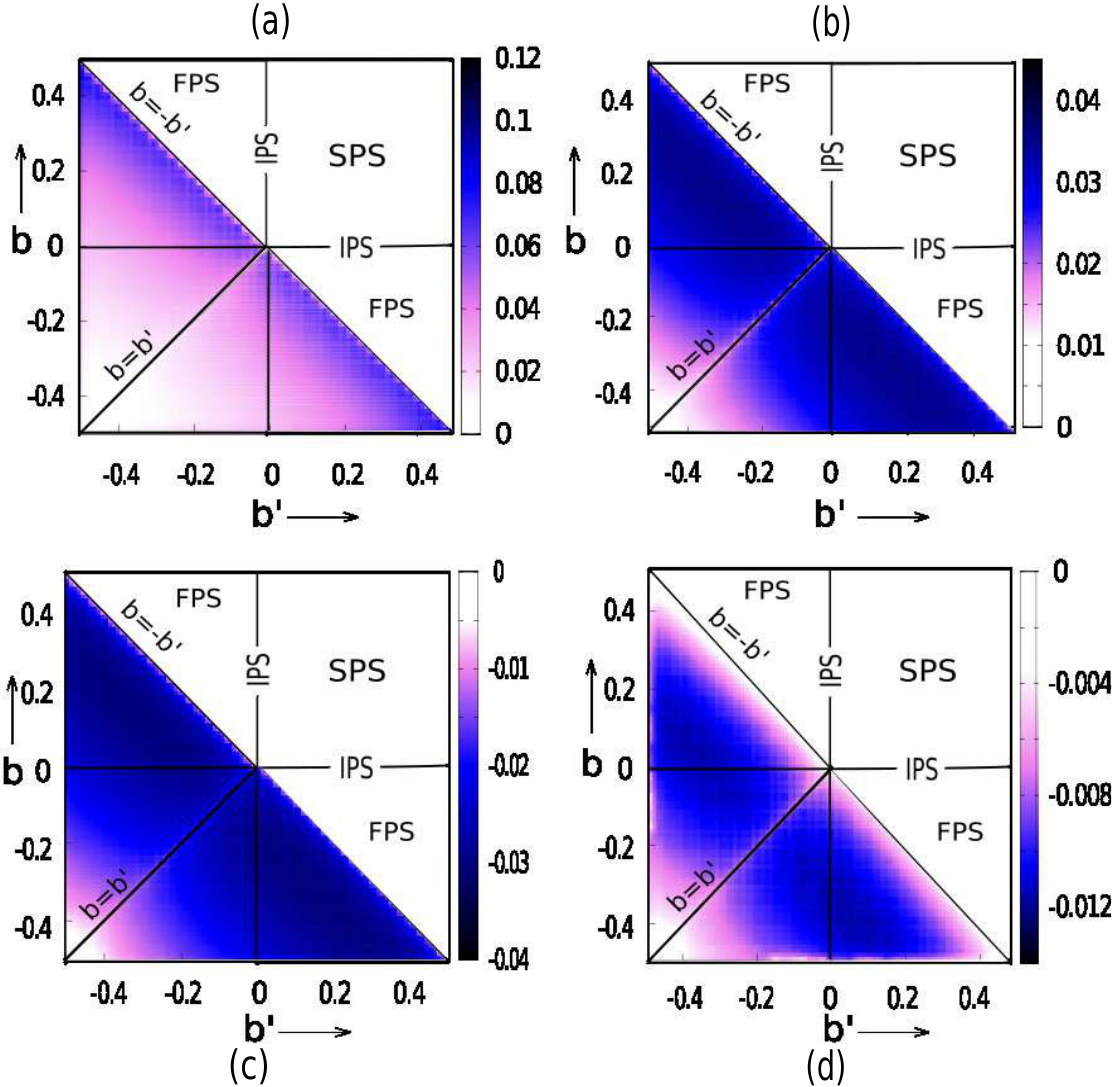} 
\caption{We plot nearest neighbor correlations in the disordered phase for $m=1/2$. The top left panel shows the data for $\langle\eta_i \eta_{i+1}\rangle-\rho^2$, the top right panel corresponds to $\langle \eta_i \sigma_i \rangle -\rho m$, the bottom left panel corresponds to $\langle \sigma_{i}\eta_{i+1} \rangle - \rho m$, and bottom right panel corresponds to $\langle\sigma_i \sigma_{i+1}\rangle-m^2$. The plots are color-coded. Out of all these four correlations, particle-particle correlations are strongest while the surface bonds show weak anticorrelations. All four correlations vanish at the $b=b'=-0.5$ point which satisfies the product measure. Here, we have used $N=2000$ and $\rho=1/2$ and all data have been averaged over $10^5$ histories.} 
\label{2pt_clrmp}
\end{figure}

In the disordered phase, the landscape and the particles show no long ranged order. However, the steady state does not satisfy product measure in general as there are short ranged correlations present in the system. In Fig. \ref{2pt_clrmp} we show the nearest neighbor correlations between the site occupancies and the bond orientations. Let $\eta_i$ be the occupancy variable for $H$ particle at site $i$, which takes the value $1$ (or $0$) if the site $i$ is occupied by an $H$ ($L$) particle. Similarly, let $\sigma_i$ denote the tilt variable which is $1$ ($0$) if the bond between sites $i$ and $i+1$ is an upslope (downslope). We measure the four nearest neighbor correlations $\langle \eta_i \eta_{i+1}\rangle$ (Fig. \ref{2pt_clrmp}, top left panel), $\langle \eta_i \sigma_{i}\rangle$ (Fig. \ref{2pt_clrmp}, top right panel), $\langle \sigma_{i-1}\eta_{i}\rangle$ (Fig. \ref{2pt_clrmp}, bottom left panel), and $\langle \sigma_i \sigma_{i+1}\rangle$ (Fig. \ref{2pt_clrmp}, bottom right panel) in steady state for different values of $b$ and $b'$ within the disordered phase.  Recall that $a >0$. From our dynamical rules in Fig. \ref{model_scheme} it follows that the model remains invariant on simultaneously interchanging $H$ $\leftrightarrow$ $L$ and $b$ $\leftrightarrow$ $b'$,  and inverting the height profiles. All correlations are therefore symmetric around the line $b=b'$ that bisects the disordered phase.  

From Fig. \ref{2pt_clrmp} we notice that $\langle \eta_i \eta_{i+1}\rangle$ correlations are strongest. Therefore, any mean-field level approximation will be affected most by this nearest neighbor correlation among the $H$ particles and we must find some parameter regime where this correlation is weak, in order for mean-field theory to work. We also notice that all four correlations are negligible near the bottom left corner of the phase diagram. In fact the corner point $b=b'=-0.5$ has been studied earlier in \cite{das01pre2} and using pairwise balance \cite{schutz96jphysa} it was shown that the system satisfies the exact product measure in this case. In the vicinity of this point, all correlations are expected to be weak and mean-field theory should work well in the neighborhood of this line.

Starting from the dynamical rules in Fig. \ref{model_scheme} we can write down the following formal expressions for the average particle current $J_\rho$ and tilt current $J_m$ in the system. 
\begin{align}
J_\rho &= (D+a)P(H \backslash L)+(D+a)P(L/H)  \nonumber \\
&-(D-a)P(L\backslash H)-(D-a)P(H/L)  \nonumber \\
J_m &=(E+b) P( /H \backslash)+(E-b^\prime)P(/L \backslash)  \nonumber \\
&-(E-b)P(\backslash H /)-(E+b^\prime)P(\backslash L /)
\label{eq:3pt_c}
\end{align}
where $P(H \backslash L)$ denotes the probability of a configuration that has an $HL$ pair in two adjacent lattice sites connected by a downslope bond ($\backslash$). Similarly, $P( /H \backslash)$ denotes the probability to have an occupied local hill. All other terms in Eqs. \ref{eq:3pt_c} may be defined in the same manner. Within the mean-field approximation, these joint probabilities can be factorized. For example, $P(H \backslash L)$ can be written as $\rho(1-m)(1-\rho)$, $P( /H \backslash)$ becomes $m\rho(1-m)$, and so on. Here, $\rho$ denotes the density of $H$ particles and $m$ denotes the density of upslope bonds in the system. The average currents can thus be written as $J_\rho = 2a \rho(1-\rho)(1-2m)$ and $J_m =2m(1-m)[\rho(b+b^\prime)-b^\prime]$. Assumption of local equilibrium means when $\rho$ and $m$ varies in space and time; local currents can still be obtained by substituting $\rho(x,t)$ and $m(x,t)$ in these expressions. We apply the formalism of NLFH illustrated in Sec. \ref{framework} starting with this expression for local currents and calculate the two mode-coupling matrices.

The phase boundary between the ordered and disordered phase can be found on noting that a ``positive feedback" (particles falling into valleys and stabilizing them) leads to the ordered phase, whereas a ``negative feedback" (particles falling into valleys, but tending to turn valleys into hills) leads to the disordered phase. The phase boundary therefore corresponds to the ``zero feedback" condition, meaning that the particles fall into valleys but do not influence the landscape dynamics (passive case). For an untilted surface, this happens when $b=-b'$, and this is then the equation of the phase boundary. Mean-field theory predicts this correctly, which can be seen as follows. Using the  mean-field expressions for $J_\rho$ and $J_m$, given in the last paragraph, we can write down the Jacobian $\mathbf A$ and its eigenvalues for $m=1/2$ are $\lambda = \pm \sqrt{-2a\rho (1-\rho) (b+b')}$, which are real for $b<-b'$ and imaginary for $b>-b'$. Imaginary eigenvalues imply that a perturbation in $\rho(x,t)$ and $m(x,t)$ grows in time and takes the system to an ordered state with macroscopic inhomogeneity. On the other hand, real eigenvalues mean traveling wave solutions hold, as discussed in Sec. \ref{framework}. Thus the $b=-b'$ line marks the boundary between ordered and disordered phases. It is remarkable that mean-field theory makes this prediction so accurately because our plots in Fig. \ref{2pt_clrmp} show that in the vicinity of the $b=-b'$ line correlations are particularly strong. We have checked that (data not shown here) for $m \neq 1/2$ the prediction does not work so well.

A somewhat improved approximation over mean-field theory would involve retaining two-point or three-point correlations in the system and factorizing the rest. For example, $P(H \backslash L)$ can be written as $P(H \backslash ) (1-\rho) $ and similarly, $P( /H \backslash) = P(H \backslash )m $, etc. Here, we have retained the correlations between a site and the next bond. These two-point correlations can be evaluated by writing down master equations for the probabilities $P( H /)$, $P(H \backslash )$, $P(L /)$ and $P(L \backslash)$ and (numerically) solving them in a self-consistent manner (see appendix \ref{app:2pt} for details).  Alternatively, one can retain three-point correlations like $P(H \backslash L)$ or $P(/H\backslash)$. These three-point probabilities can again be evaluated by writing down the corresponding master equations and solving for steady state (details in appendix \ref{app:3pt}). We compare the current $J_\rho$ and $J_m$ as well as eigenvalues of the Jacobian matrix obtained from different approximation schemes and simulations for a few representative values of $b,b',\rho,m$ in Tables \ref{tab1} and \ref{tab2} in Appendix \ref{app:tab}.

However, our final conclusions are not so sensitive to whether we neglect all correlations in the system as in mean-field theory, or include two or three-point correlations in our description. Using the NLFH method, when we calculate the mode-coupling matrices ${\mathbf G}^1$ and ${\mathbf G}^2$, the condition of observing various  universality classes depends on whether certain matrix elements are zero or nonzero. The actual value of these matrix elements may differ depending on the approximations used, but that does not change the dynamical universality class. We carry out our analysis within that region of the disordered phase, where correlations are weak (see Fig. \ref{2pt_clrmp}) and thus find no significant difference (results not shown here) based on our approximation scheme.


\section{Simulation results for dynamical structure function}
\label{dynamics_kw}
As mentioned in Sec. \ref{intro}, our model is rich enough to show many different dynamical universality classes in different parts of the disordered phase shown in the phase diagram in Fig. \ref{phase_d}. In this section, we demonstrate this by measuring the dynamical structure function $C_{\alpha \alpha}(x,t) = \langle \phi_\alpha (0,0) \phi_\alpha (x,t) \rangle $ with $\alpha =1,2$ in simulations. The angular brackets denote the average over the steady state ensemble. We extract the value of $\lambda_\alpha$ (see Eq. \ref{eq:stf_real}) from simulation and find out which value of the dynamical exponent $z_\alpha$ gives the best scaling collapse. To test the predictions from NLFH, we compare this $z_\alpha$ with the value obtained from our NLFH calculations. We find finite size effects can significantly affect the estimate of $z_\alpha$. We first demonstrate this for the point $b=b'=-0.5$, where the product measure condition is valid and exact expressions for $J_\rho$ and $J_m$ are available \cite{das01pre2}.

\subsection{Significant finite size effects for $b=b'=-0.5$} 
\label{sec:prod}
\begin{figure}[h!]
\begin{center}
\includegraphics[scale=0.55]{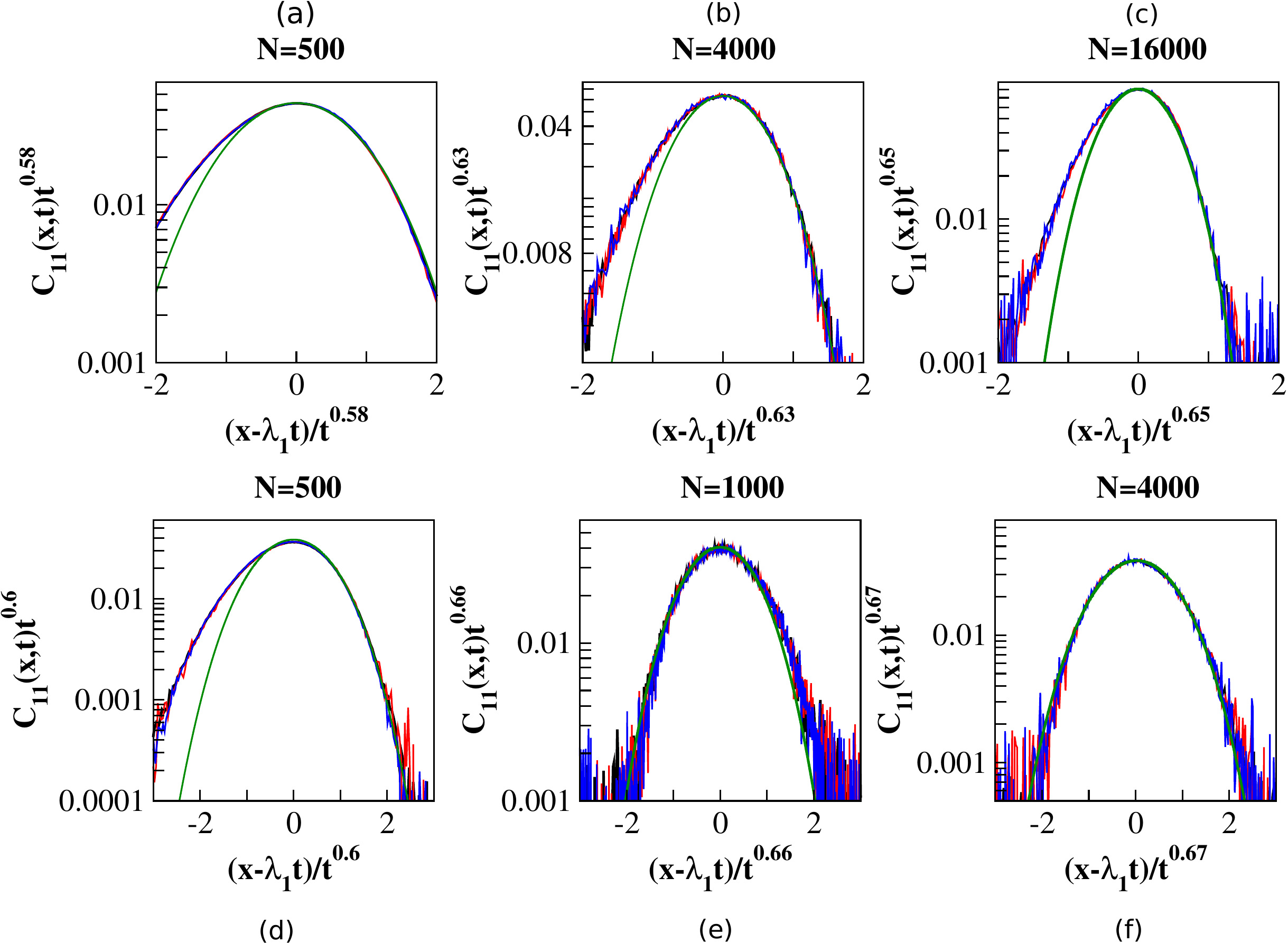}
\caption{Scaling collapse of dynamical structure function $C_{11}(x,t)$ 
for which a KPZ mode is expected, but for smaller system size  significant deviation is observed. The upper panel corresponds to  $b=-0.5,b^\prime=-0.5,\rho=0.3,m=0.5$ for which $G^1_{11}=G^2_{22}=-0.6$. We find strong finite size effects in this case. Even for $N=16000$ the best collapse is obtained for $1/z=0.65$ and the scaling function also shows significant deviation from the Pr\"ahofer-Spohn function (green line). For smaller $N$ the effect is even stronger. The bottom panel corresponds to $b=-0.5,b^\prime=-0.5,\rho=0.8,m=0.6$ for which $G^1_{11}\simeq -0.89$ and $G^2_{22} \simeq -0.51$. In this case finite size effects are much weaker. Since the product measure holds for  $b=-0.5,b^\prime=-0.5$, the $G$-matrix elements are obtained from mean-field theory which is exact in this case. All data have been averaged over at least $10^5$ histories. The error bar in $1/z$ is $\pm0.005$ in the left panel, while in the middle and right panels, the error bar is $\pm 0.01$. }
\label{kpz_stf_fse_merged}
\end{center}
\end{figure}

For the product measure point, earlier studies have shown that for different values of $\rho$ and $m$ the system can show KPZ and diffusive modes \cite{das01pre2}. Consider the case for two KPZ modes, when $G^1_{11}$ and $G^2_{22}$ both are nonzero. In this case, we expect a dynamical exponent $z_\alpha=3/2$ and $(x-\lambda_\alpha t)/t^{2/3}$ to be the scaling variable. In the top panel of Fig. \ref{kpz_stf_fse_merged} we present data for a particular set of values for $\rho$ and $m$ for which $G^1_{11}$ and $G^2_{22}$ are nonzero but their values are not so large, $G^1_{11}=G^2_{22} = -0.6$. Plots \ref{kpz_stf_fse_merged}(a),(b) and (c) show the scaling collapse for $C_{11}(x,t)$ for different system sizes. We find strong finite size effects in the scaling collapse. Even for the largest possible system size we could access ($N=16000$), we find deviation from KPZ scaling, and for smaller $N$ the deviation is even larger. On the other hand, in Fig. \ref{kpz_stf_fse_merged}(d),(e) and (f) we show the scaling collapse for $C_{11}(x,t)$ for another set of $\rho,m$ values, for which $G^1_{11}=-0.89$, $G^2_{22}=-0.51$. The self-coupling term for the first mode is now larger than before and in this case we find much weaker finite size effect: for $N=4000$ good agreement with KPZ exponent is obtained for the first mode. Our data in Fig. \ref{kpz_stf_fse_merged} also show that for smaller $N$ values, the shape of the master curve is not completely symmetric and the left tail is slightly longer than the right tail. However, as $N$ becomes larger the symmetry is restored, as expected for the Pr\"ahofer-Spohn scaling function \cite{prahofer04jsp}.

It is easy to see why finite size effects are stronger for smaller values of $G^\alpha_{\alpha \alpha}$. In Eq. \ref{eq:scaling_fn} since $Q_{\alpha \alpha}$ is proportional to $(G^\alpha_{\alpha \alpha})^2$, when the self-coupling co-efficient $G^\alpha_{\alpha \alpha}$ has a small value, the third term on the right hand side of this equation also becomes small. In the limit $k \to 0$, this third term alone is expected to survive and the other terms should vanish. However, for finite system size $N$ the smallest possible value of $k$ is $2 \pi /N$ and it is possible that if $N$ is not so large the diffusive and cross-coupling terms become comparable to the self-coupling term and affect the apparent value of $z_\alpha$ and the nature of the scaling function.  For example, in the case when we expect two KPZ modes, $z_\alpha = z_\beta=3/2$, it follows from Eq. \ref{eq:scaling_fn} that the diffusive term vanishes as $N^{-1/2}$ and the cross-coupling term scales as $N^{-1/6}$. Due to such slow decay, one really needs to consider very large values of $N$ such that $k$ is small enough for the effect of the diffusive and cross-coupling terms in Eq. \ref{eq:scaling_fn} to be ignored. For larger value of $G^\alpha_{\alpha \alpha}$, the self-coupling term is already large, and the diffusive and cross-coupling terms are relatively small even when $N$ is not so large.

Thus we see that although NLFH predicts a KPZ universality class for nonzero self-coupling, in order to numerically observe the same, it is not sufficient that $G^\alpha_{\alpha \alpha}$ is nonzero; it should also have a sufficiently large value. Otherwise, finite size effects can become very strong and the value of the dynamical exponent, as well as the nature of the scaling function may be significantly affected.
Note that within our model, the magnitude of $G^\alpha_{\alpha \alpha}$  cannot be arbitrarily large and the upper bound, estimated from mean-field theory is  $\sim 2$. We find that $G^\alpha_{\alpha \alpha} \gtrsim 0.9$ can be considered to be sufficiently large and yields good KPZ scaling and if  $G^\alpha_{\alpha \alpha}$ falls below $\sim 0.5$, we do not find good scaling for the largest possible system size $N=16000$ accessible to us.  Our data in Fig. \ref{kpz_stf_fse_merged} are for the parameter values where the product measure holds and the exact expressions for currents are available. But this issue becomes even more crucial when currents are not exactly known and approximate expressions are used in NLFH analysis. In that case we have to rely more heavily on numerics and it then becomes even more important that our numerical observation of the scaling collapse is not plagued by finite size effects. In the following subsections we show a few such examples. Unless otherwise mentioned, in all our data for the dynamical structure factor below, we have used $N=16000$. 


\subsection{KPZ and $5/3$ L\'evy mode}
\label{sub_53l}
As discussed in Sec. \ref{framework}, the condition for having mode $\alpha$ in the KPZ universality class and mode $\beta$ in the $5/3$ L\'evy class is 
\begin{eqnarray}
G^\alpha_{\alpha \alpha},G^\beta_{\alpha \alpha} \neq 0, G^\beta_{\beta \beta}=0.
\nonumber \\ \label{eq:53l_1}
\end{eqnarray}
Finite size corrections play an important role here too. For KPZ scaling the self-coupling term in Eq. \ref{eq:scaling_fn} survives in the $k \rightarrow 0$ limit while the diffusive term is $\sim k^{1/2}$ and the cross-coupling term is $\sim k^{1/10}$.  On the other hand, for $5/3$ L\'evy scaling, the cross-coupling term survives in the small $k$ limit, the self-coupling co-efficient $G^\beta_{\beta \beta}$ vanishes and the  diffusive term is $\sim k^{1/3}$. Therefore, both $G^\alpha_{\alpha \alpha}$ and $G^\beta_{\alpha \alpha}$ should have large magnitudes in order to quell these strong finite size effects. Across all values of the parameters $b,b',\rho,m$ the magnitude of $G^\beta_{\alpha \alpha}$ shows an upper bound $\sim 1.8$ while $G^\alpha_{\alpha \alpha}$ stays below $\sim 2$ (also see the previous section). We observe numerically that $G^\alpha_{\alpha \alpha}$ should be at least as large as $\sim 0.6$ and $G^\beta_{\alpha \alpha} \gtrsim 0.35 $ for $N=16000$. The other coefficient $G^\beta_{\beta \beta} \lesssim 0.1$ is found to be good enough for our purpose.

%
%

\begin{figure}[h!]
\begin{center}
\includegraphics[scale=1.5]{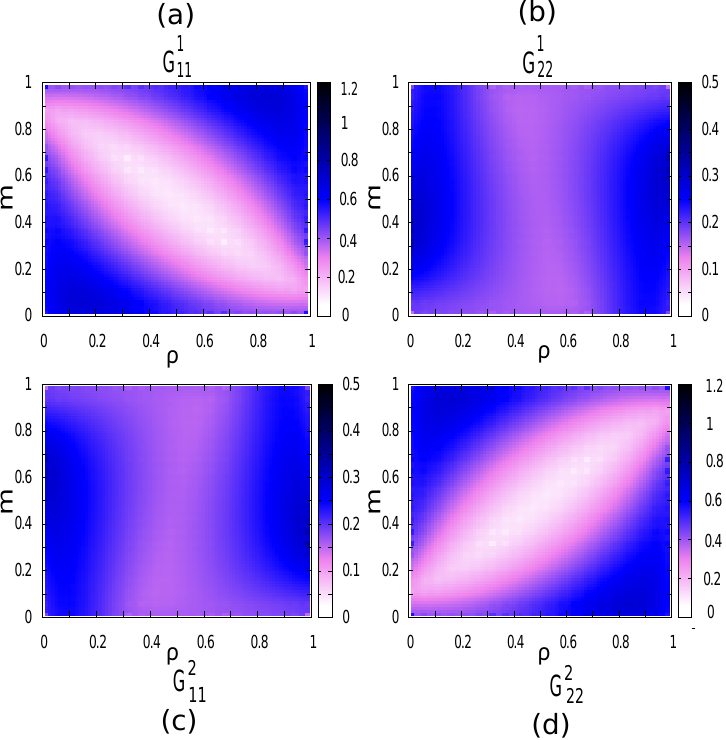}
\caption{Contour plot of all the four diagonal elements of the two mode-coupling matrices in $\rho$-$m$ plane for $b=-0.3,b^\prime=-0.3$. The color coding is presented next to each plot. Diagonal panels show the self-coupling terms and the off-diagonal panels show the cross-coupling terms. From these maps it is possible to find a few regions in the $\rho$-$m$ plane where the $5/3$-L\' evy mode is expected, according to the criterion in Eq. \ref{eq:53l_1}. We have used mean-field theory to obtain these plots.}
\label{53L_clrmp}
\end{center}
\end{figure}

We choose $b=b'=-0.3$, a point where spatial correlations are expected to be weak (see Fig. \ref{2pt_clrmp}) and our approximate expressions for $J_\rho$ and $J_m$ agree reasonably well with currents measured in simulations (comparison not shown here). For this particular $b$ and $b'$ we plot the values of the diagonal elements of ${\mathbf G}^1$ and ${\mathbf G}^2$ in the $\rho$-$m$ plane in Fig. \ref{53L_clrmp}.   

\begin{figure}[ht!]
\begin{center}
\includegraphics[scale=0.8]{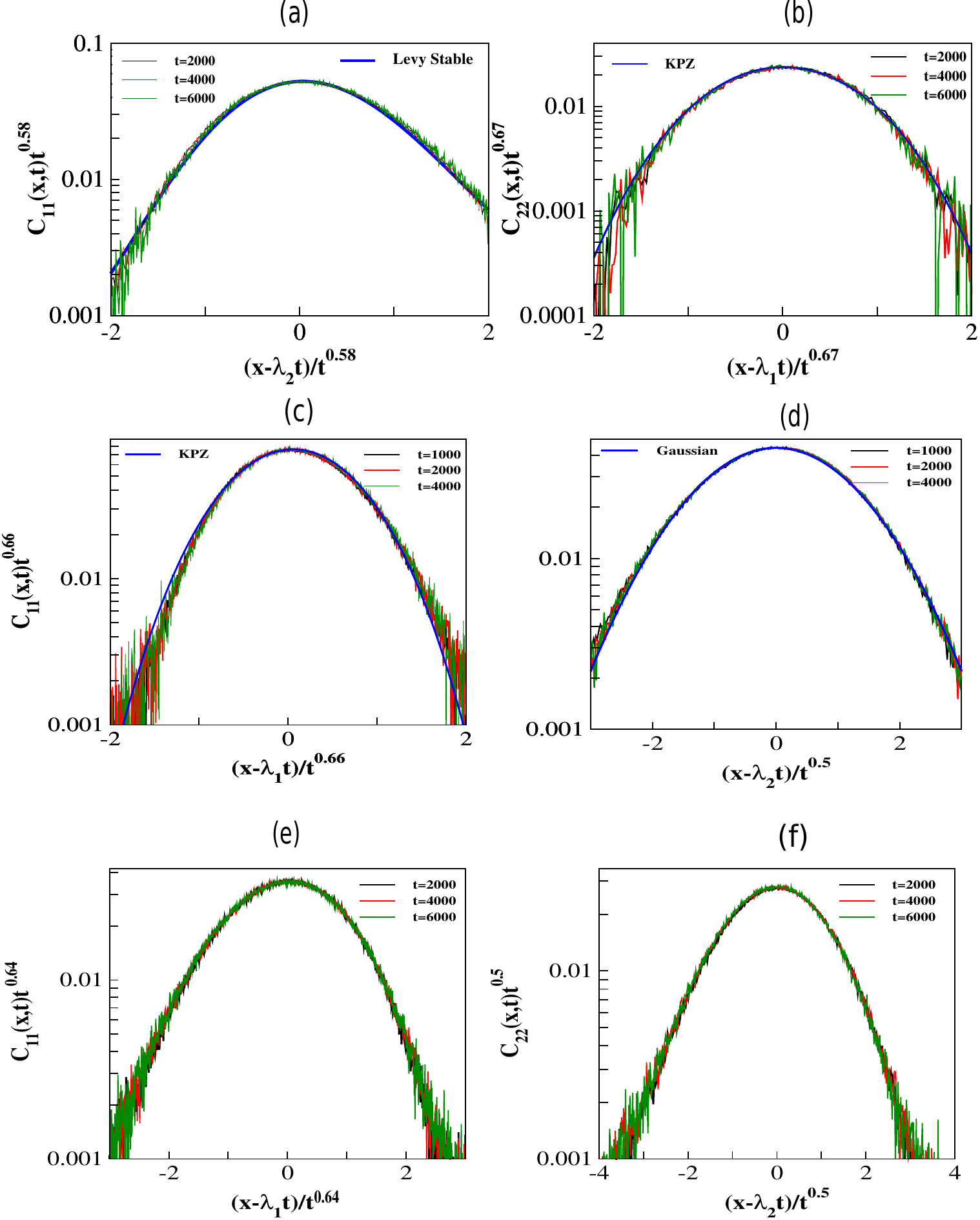}
\caption{Numerical verification of $5/3$ L\'evy and KPZ modes for $b=b'=-0.3$. The top panel is for $\rho=0.89$ and $m=0.23$ and for $G^1_{11}=-0.06,G^1_{22}=-0.39,G^2_{22}=-1.01$. (a) shows a scaling collapse with $1/z_1 = 0.58$ which is close to the value $0.6$ expected for the $5/3$ L\'evy universality class. The scaling function fits well with $\alpha$-stable distribution with $\alpha = 1.72$. (b) shows KPZ scaling with good fit with the Pr\"ahofer-Spohn scaling function. The middle panel is for $\rho=0.31,m=0.32$ for which $G$-matrix values are  $G^1_{11}=0.512,G^2_{11}=0.126,G^2_{22}=-0.003$. (c) shows good agreement with KPZ scaling, as expected. The small value of $G^2_{11}$, however, gives rise to strong finite size effects which masks the $5/3$ L\'evy class and shows diffusive scaling instead in (d). The bottom panel is for $\rho=0.915,m=0.875$. Here, $G^1_{11}=-1.07,G^2_{11}=-0.431,G^2_{22}=-0.002$ and one would expect first mode KPZ and second mode $5/3$ L\'evy. However, (e) and (f) show significant deviation from both these values. For all cases, the $G$-matrix elements are calculated by retaining two-point correlations between a site and the next bond and ignoring the rest. All data have been averaged over at least $10^5$ independent histories. The error bar in $1/z$ values is $\pm 0.01$.}
\label{53L_kpz_stf_merged}
\end{center}
\end{figure}

From this plot, we see that in the bottom-right region in the $\rho$-$m$ plane, $G^1_{11}$ has a small value, $G^2_{22}$ is large, and $G^1_{22}$ is also moderately large which makes this region the best choice for observing the $5/3$ L\'evy universality class for mode $1$ and KPZ class for mode $2$. We present our data in Fig. \ref{53L_kpz_stf_merged}(a),(b) for $\rho=0.89$ and $m=0.23$. In plot \ref{53L_kpz_stf_merged}(a), we show our simulation data for $C_{11}(x,t)$ and find the best collapse is obtained when the shifted $x$ axis is rescaled with $t^{0.58}$, which is close to the value $3/5$ expected in this case. We also compare the master curve with  the $\alpha$ L\'evy stable distribution where $\alpha = 1/0.58=1.72$ and find quite a good fit. In Fig. \ref{53L_kpz_stf_merged}(b) we show the scaling collapse for $C_{22}(x,t)$ and in this case we observe a dynamical exponent $z_2=3/2$ and our master curve also matches well with the Pr\"ahofer-Spohn scaling function \cite{prahofer04jsp}. 

To demonstrate the finite size effect in this case, we now choose another point in the $\rho$-$m$ plane, $\rho=0.31$ and $m=0.32$. From Fig. \ref{53L_clrmp} we can see that at this point $G^1_{11}$ is large, $G^2_{22}$ is almost zero but $G^2_{11}$ is small. These values are such that while mode $1$ is expected to show good agreement with the KPZ universality class, the observation of the $5/3$ L\'evy universality class for mode $2$ may not be possible due to finite size effects. Indeed our data in Fig. \ref{53L_kpz_stf_merged}(c),(d) show that the scaling collapse for $C_{11}(x,t)$ has been obtained for $1/z_1 \simeq 0.66$, which is  close to the KPZ exponent, but $C_{22}(x,t)$ shows a scaling collapse with effective $z_2 \simeq 2$ which corresponds to the diffusive universality class, instead of $z_2=5/3$. Even the scaling function in this case matches well with a Gaussian function which is the scaling function expected for a diffusive mode. In other words, the self-coupling term for mode $2$ being close to zero in this case, in Eq. \ref{eq:scaling_fn} only the diffusive term and the cross-coupling term are present and due to the small magnitude of the cross-coupling term, the diffusion term dominates the scaling behavior. Our choice of $N=16000$ is not large enough to remove this strong finite size effect and it is not numerically feasible to consider $N$ much larger than this.

We present a third scenario for a KPZ and $5/3$ L\'evy combination, where we choose a point in the $\rho$-$m$ plane which is close to one corner such that both $\rho$ and $m$ are high or low. We find in this case, although $G^1_{11}$ and $G^2_{11}$ are significantly large and $G^2_{22}$ is negligibly small, we do not find KPZ and $5/3$ L\'evy universality classes. In Fig. \ref{53L_kpz_stf_merged}(e),(f) we show our data. We do not yet have any explanation for this result.

\subsection{Modified KPZ and diffusive mode}
\label{sub_modkpz}
\begin{figure}[h!] 
\begin{center} 
\includegraphics[scale=0.6]{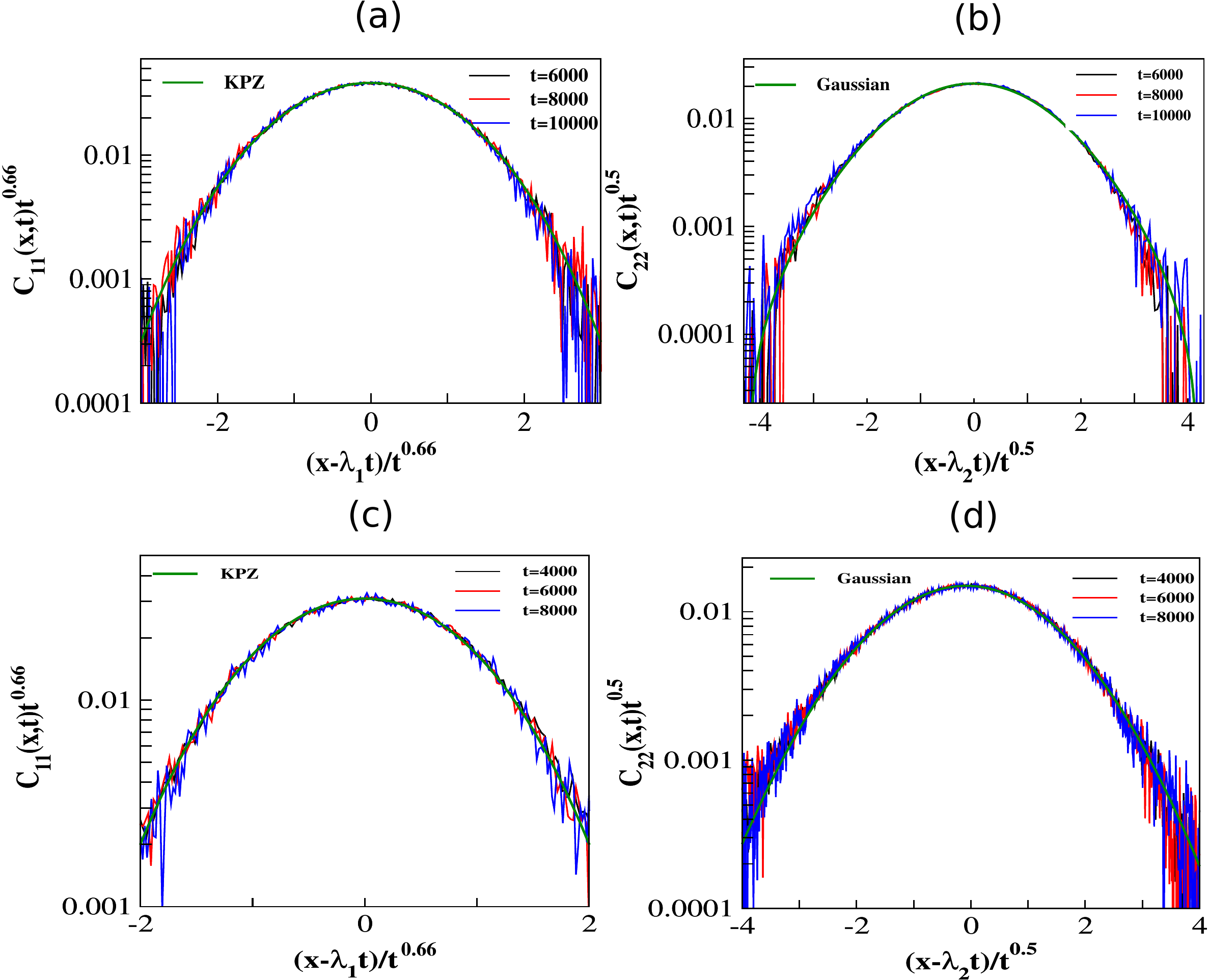}
\caption{Numerical verification of modified KPZ and diffusive modes. The top panel uses $b=-0.5,b^\prime=-0.5,\rho=0.92,m=0.93$. Corresponding $G$-matrix elements are exactly known at these parameter values since the product measure holds. These elements are $G^1_{11}=1.8,G^1_{22}=0.6,G^2_{11}=0.025,G^2_{22}=-0.076$.   Although $C_{11}(x,t)$ is expected to show modified KPZ scaling here, the scaling function in (a) fits well with usual KPZ scaling function. (b) shows diffusive scaling as expected. The bottom panel is for $b=-0.3,b^\prime=-0.5, \rho=0.12, m=0.105$. The $G$-matrix elements in this case are are obtained from currents which are calculated in a self-consistent manner by retaining two-point correlations between a site and the next bond and ignoring all other correlations in the system. The values obtained thus are: $ G^1_{11} = 1.52, G^1_{22} =-0.501, G^2_{11} = 0.023, G^2_{22} = 0.004$. Even in this case (c) shows usual KPZ scaling, rather than modified KPZ. Diffusive scaling for $C_{22}(x,t)$ is shown in (d). The error bar in the $1/z$ value for (c) is $\pm 0.02$ and for (a), (b), and (d) is $\pm 0.01$. An averaging over at least $10^5$ histories has been performed.} \label{modkpz_merged} 
\end{center} 
\end{figure}

The criterion for observing the $\alpha$ mode in a modified KPZ class and $\beta$ mode in a diffusive class is 
\be 
G^\alpha_{\alpha \alpha}, G^\alpha_{\beta \beta} \neq 0, G^\beta_{\beta \beta} = G^\beta_{\alpha \alpha} =0  \label{eq:modkpz_1}
\ee
This criterion can be satisfied for different set of $b,b'$ values. First we present our data for $b=b'=-0.5$ where the product measure holds and the exact expression for currents is available \cite{das01pre2}. In Fig. \ref{modkpz_merged}(a), (b) we present our simulation data for one particular choice of $\rho$ and $m$. From the mode-coupling matrix elements given in the figure caption, it is clear that for the first mode we expect a modified KPZ behavior, while for the second mode we expect diffusive scaling. Figure \ref{modkpz_merged}(a) shows the structure function for mode $1$ which shows a good scaling collapse for $1/z_1 = 0.66$, which is close to the expected value $2/3$. Interestingly, our master curve fits rather well with the usual Pr\"ahofer-Spohn scaling function. Figure  \ref{modkpz_merged}(b) shows the structure function of the second mode and as expected, it belongs to the diffusive universality class. 

\begin{figure}[h!]
\begin{center}
\includegraphics[scale=1.5]{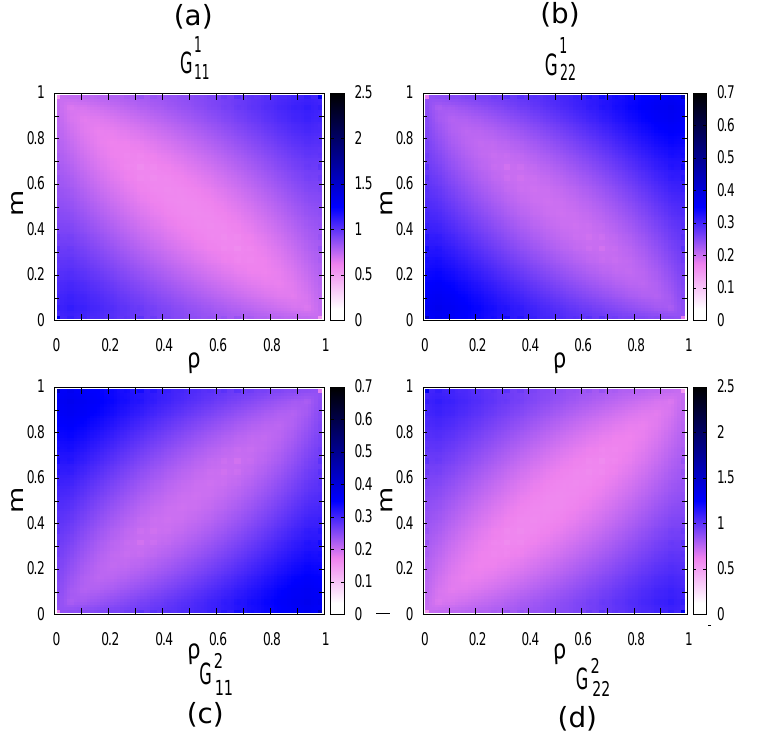}
\caption{Contour plots for four diagonal elements of the two $G$ matrices in the $\rho$-$m$ plane for $b=-0.3,b^\prime=-0.5$. In various different regions in the $\rho$-$m$ plane the criterion for modified KPZ and diffusive modes as given in Eq. \ref{eq:modkpz_1} is satisfied. These plots are obtained using mean-field expressions for currents.}
\label{modkpz_clrmp}
\end{center}
\end{figure}

To probe further the observed similarity between the modified KPZ and usual KPZ scaling function, we examine the specific values of the mode-coupling matrix elements. From the caption of Fig. \ref{modkpz_merged} we notice that the self-coupling term $G^1_{11}$ is almost three times larger than the cross-coupling term $G^1_{22}$. So it is possible that the cross-coupling is not felt so strongly and the mode shows usual KPZ scaling. We have extensively searched in our parameter space but could not find any $(b,b',\rho, m)$ set for which Eq. \ref{eq:modkpz_1} is satisfied, and $G^\alpha_{\alpha \alpha}$ is smaller than $ G^\alpha_{\beta \beta}$.  We observe that $G^\alpha_{\alpha \alpha} \gtrsim 0.8$ and $ G^\alpha_{\beta \beta} \gtrsim 0.5$ in all cases where we have spotted modified KPZ and diffusive universality classes. We show one example in Fig. \ref{modkpz_clrmp}, where we plot the mode-coupling matrix elements in the $\rho$-$m$ plane for one specific $(b,b')$ set. Although Eq. \ref{eq:modkpz_1} is satisfied for many $(\rho,m)$ values, for each of them we find the self-coupling term is significantly larger than the cross-coupling term. Our simulation data shows usual KPZ scaling in this case also (see Fig. \ref{modkpz_merged}c). Note that the exact scaling function for the modified KPZ universality class is not known and the observed similarity with usual Pr\"ahofer-Spohn scaling function may also indicate that the two scaling functions are actually the same.  


\subsection{Golden mean modes}
\label{sub_gm}
\begin{figure}[ht!]
\begin{center}
\includegraphics[scale=1.3]{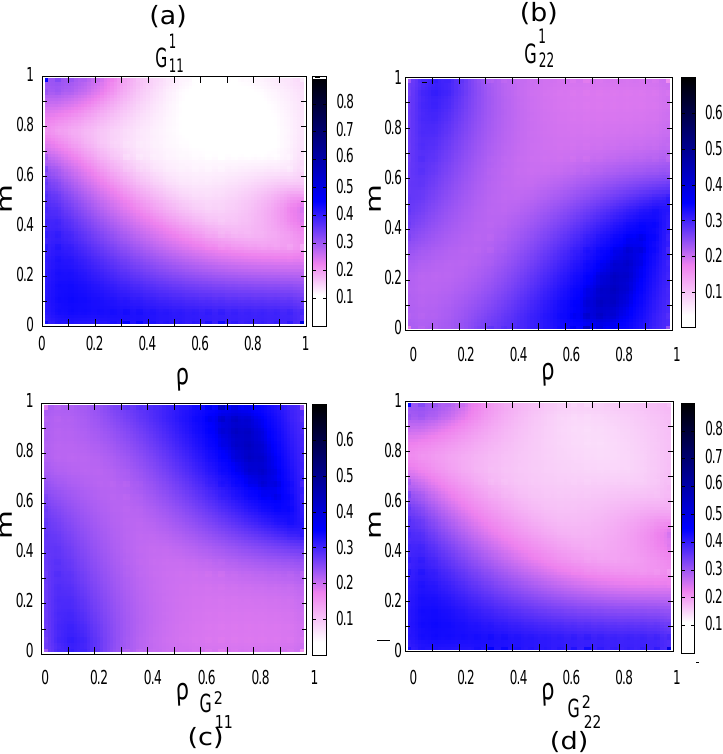}
\caption{A contour plot for four diagonal elements of the two mode-coupling matrices in the $\rho$-$m$ plane for $b=0.1,b^\prime=-0.3$. These color-coded plots show that in no region of the $\rho$-$m$ plane do we have simultaneous vanishing of self-coupling terms and large cross-coupling terms. Hence the criterion in Eq. \ref{gm_eq} for golden mean modes is never satisfied. Similarly, it follows that the condition for the $3/2$ L\'evy and diffusive mode pair, given in Eq. \ref{eq:32l}, is also not met. These plots are obtained using mean-field expressions for currents.}
\label{gm_clrmp}
\end{center}
\end{figure}

\begin{figure}[ht!]
\begin{center}
\includegraphics[scale=0.6]{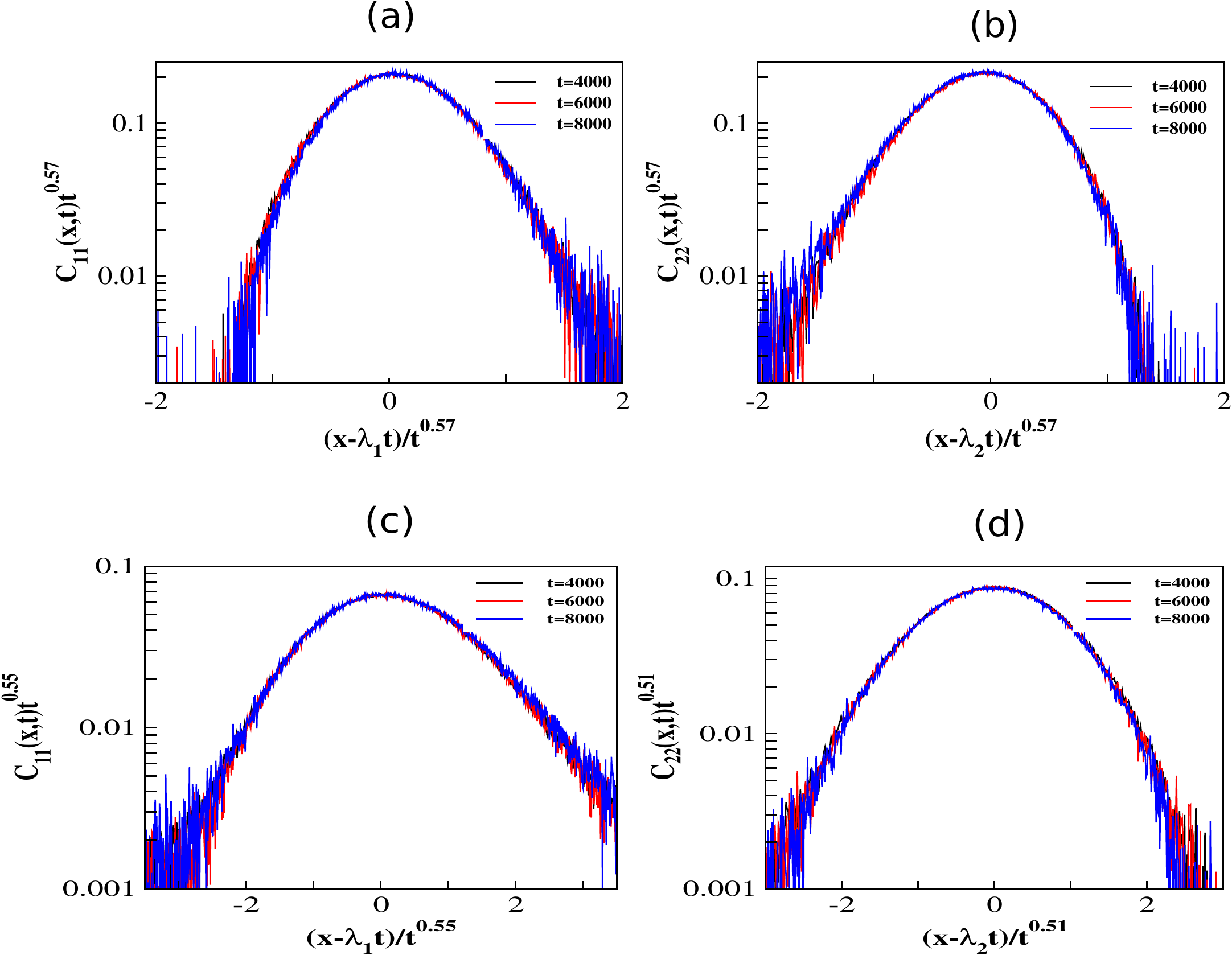}
\caption{The top panel shows the scaling collapse for the dynamical structure factor for $b=0.1,b^\prime=-0.3,\rho=0.67,m=0.49$. The $G$-matrix elements in this case are obtained from currents which are calculated in a self-consistent manner by retaining three-point correlations between objects consisting of two sites and the intermediate bond or two bonds and an intermediate site and ignoring all other correlations in the system. The values obtained thus are
$G^1_{11}=-0.072,G^1_{22}=0.228,G^2_{11}=-0.179,G^2_{22}=0.026$. Although golden mean modes are expected theoretically, the cross-coupling terms being not so large, finite size effects significantly change the value of the dynamical exponent. Instead of $1/z \simeq 0.618$ we find here an exponent $0.57 \pm 0.02$ for both the modes. The bottom panel is for the same $b,b'$ values but $\rho=0.85,m=0.34$. Here, $G^1_{11}=0.105,G^1_{22}=0.72,G^2_{11}=-0.154,G^2_{22}=-0.027$. Instead of $3/2$ L\'evy scaling for the first mode, we observe $1/z=0.55\pm0.01$. (d) shows a diffusive scaling for the second mode as expected. These data have been averaged over at least $10^5$ independent histories.}
\label{gm_levy32_merged}
\end{center}
\end{figure}

Golden mean modes always occur in pairs, because the dynamical exponents $z_\alpha$ and $z_\beta$ satisfy the conditions $z_\alpha = 1+1/z_\beta$ and $z_\beta = 1+1/z_\alpha$, the recursive solution of which yields $z_\alpha = z_\beta = (\sqrt{5}+1)/2$. In our system there are only two modes and hence both $C_{11}(x,t)$ and $C_{22}(x,t)$ should show scaling as per the golden mean universality class. This happens when the self-coupling term vanishes and the cross-coupling term survives for each mode:  
\begin{eqnarray}
G^1_{22},G^2_{11} \neq 0; G^1_{11} = G^2_{22}=0 \label{gm_eq}
\end{eqnarray} 
However, we find that in our system these criteria are not simultaneously satisfied. We could not find any point in our parameter space where both cross-coupling terms are sufficiently large (to avoid finite size effects) and self-coupling terms are negligibly small. We illustrate this in Fig. \ref{gm_clrmp} where we have shown the variation of these matrix elements in the $\rho$-$m$ plane for a fixed $b$ and $b'$. The top right and bottom left panels show the variation of the cross-coupling co-efficients and it is clear from the color shades in these two panels that whenever one cross-coupling term gets large, the other one becomes small. Therefore, the condition in Eq. \ref{gm_eq} is not satisfied. Our simulation results for the structure functions confirm this reasoning. We have determined the dynamical exponents in this case by measuring the variance of the structure functions as a function of time (data not shown) and found that for both modes, the variance scales as $t^{0.57}$, whereas for the golden mean an exponent $\simeq 0.62$ should be obtained. 
In Fig. \ref{gm_levy32_merged}(a), (b) we show the scaling collapse of $C_{11}(x,t)$ and $C_{22}(x,t)$; both modes show good collapse with $\sim t^{0.57}$ scaling. 

\subsection{$3/2$ L\'evy and diffusive mode}
\label{sub_32l}
The criterion for observing a $3/2$ L\'evy universality class for the mode $\alpha$ and diffusive class for mode $\beta$ is
\be 
G^\alpha_{\alpha \alpha} = G^\beta_{\alpha \alpha} = G^\beta_{\beta \beta }=0, G^\alpha_{\beta \beta } \neq 0. \label{eq:32l}
\ee
We find that in our system this criterion is not satisfied for any parameter regime. Although it is possible to find the self-coupling term for both modes and the cross-coupling term for the mode $\beta$ simultaneously small, the cross-coupling term for mode $\alpha$ also tends to be small in this case. As a result, we are not able to observe  the
$3/2$ L\'evy mode in our system. We show one example in Fig. \ref{gm_levy32_merged}(c),(d).

\section{Conclusion}
\label{con}
In this paper, we have studied  the dynamics in the disordered state of a coupled system of sliding particles on a fluctuating landscape using the recently developed formalism of NLFH. In most of our parameter space, the product measure does not hold and thus the exact current-density relationship is not known. We restrict our paper to those regions of the parameter space where spatial correlations are weak and use the mean-field approximation and also improved approximations within which short ranged correlations are calculated self-consistently. Using the resulting approximate expressions for currents we perform an NLFH calculation which predicts the existence of $5/3$ L\' evy, $3/2$ L\' evy, golden mean and modified KPZ universality classes, apart from the usual KPZ and diffusive classes. However, when we attempt to test these predictions from our numerical simulations, we encounter strong finite size effects. Eq. \ref{eq:scaling_fn} predicts that subleading corrections can fall extremely slowly with system size $N$, for example, as $N^{-1/10}$ in the case of KPZ scaling when the other mode shows $5/3$ L\'evy scaling, and as $N^{-1/6}$ when there are two coupled KPZ modes. It follows that it is not enough to have a certain mode-coupling coefficient be nonzero; its magnitude needs to be large enough for that term to dominate in the numerically accessible range of system size. This makes it difficult for us to observe the golden mean or $3/2$ L\' evy universality classes in our system. However, we have been able to verify the existence of the $5/3$ L\' evy universality class. The case of the modified KPZ universality class yields an interesting outcome. Although this universality class is characterized by dynamical exponent $3/2$, its scaling function is thought to be different from the Pr\"ahofer-Spohn function \cite{popkov16jsm}. However, our data show that the master curve obtained after scaling collapse fits the Pr\"ahofer-Spohn function quite well. This is true even for  $b=b'=-0.5$, where the product measure holds and exact expressions for currents are known \cite{das01pre2}. Since the exact form of the modified KPZ scaling function has not been calculated yet, one cannot rule out the possibility that it coincides with or is extremely close to the KPZ scaling function.

In conclusion, in the study of driven diffusive systems, it is important to extend the formalism of NLFH for systems where the exact measure is not known. Our paper takes a step in that direction and shows how finite size effects can sometimes overshadow the predictions of NLFH. It would be interesting to have a general understanding of the importance of finite size effects for various different unconventional universality  classes. We hope our paper will encourage more activity in this direction.

\section{Acknowledgements} 
We acknowledge useful discussions with G.M. Sch\"utz, H. Spohn, and S. Mahapatra. This research was supported in part by the International Centre for Theoretical Sciences (ICTS) during a visit for participating in the program Universality in Random Structures: Interfaces, Matrices, Sandpiles (Code No. ICTS/urs2019/01) and the program Indian Statistical Physics Community Meeting (Code No. ICTS/ispcm2019/02). S.C. acknowledges financial support from the Science and Engineering Research Board, India (Grant No. EMR/2016/001663). The computational facility used in this work was provided through the Thematic Unit of Excellence on Computational Materials Science, funded by Nanomission, Department of Science and Technology (India).

\appendix
\section{Time evolution equation for two-point correlation functions}
\label{app:2pt} 
In this appendix, we present the time-evolution equation for the two-point correlation functions between the occupancy of a site and the tilt of the bond on its right. We denote these correlators as $P(H /)$, $P(H \backslash)$, $P(L/)$ and $P(L \backslash)$, where $P(H /)$ denotes the probability to find an $H$ particle and an upslope bond next to it. The other three quantities can also be explained similarly. The time-evolution equations can be written as
\begin{align}
\frac{dP(H/)}{dt} &=(E-b)P(H \backslash)P(H/)+(E+b^\prime)P(H \backslash)P(L/) \nonumber \\ 
&+(E+b)P(H \backslash)m+P(H \backslash)P(L/)+P(L/)\rho-(E+b)P(H/)P(H \backslash) \nonumber \\
&-(E-b^\prime)P(H/)P(L \backslash)-(E-b)P(H/)(1-m)-P(L/)P(H/) \label{eq:a1} \\
\frac{dP(H \backslash)}{dt} &= (E-b)P(H/)(1-m)+(E+b)P(H/)P(H \backslash) \nonumber \\
&+(E-b^\prime)P(H/)P(L \backslash)+P(H \backslash)P(L \backslash)-(E+b)P(H \backslash)m \nonumber \\
&-(E-b)P(H \backslash)P(H/)-(E+b^\prime)P(H \backslash)P(L/)-P(L/)P(H \backslash)\nonumber \\
&-P(H \backslash)(1-\rho) \\
\frac{dP(L/)}{dt} &=P(L/)P(H/)+(E-b^\prime)P(L \backslash)m+(E-b)P(L \backslash)P(H/) \nonumber \\
&+(E+b^\prime)P(L \backslash)P(L/)-P(L/)\rho-P(H \backslash)P(L/)-(E+b)P(L/)P(H \backslash) \nonumber \\
&-(E-b^\prime)P(L/)P(L \backslash)-(E+b^\prime)P(L/)(1-m) \\
\frac{dP(L \backslash)}{dt} &=(E+b^\prime)P(L/)(1-m)+(E+b)P(L/)P(H \backslash) \nonumber \\
&+(E-b^\prime)P(L/)P(L \backslash)+P(H \backslash)(1-\rho)+P(L/)P(H \backslash) \nonumber \\
&-(E-b^\prime)P(L \backslash)m-(E-b)P(L \backslash)P(H/)-(E+b^\prime)P(L \backslash)P(L/) \nonumber \\
&-P(H \backslash)P(L \backslash)
\label{eq:2pt_me}
\end{align}
where the first term on the right hand side of Eq. \ref{eq:a1} corresponds to the case when a local configuration of the form $H \backslash H / $ changes to $H / H \backslash$ with the rate $(E-b)$. If the two $H$ particles here are assumed to be at sites $i$ and $i+1$, then this transition increases the probability of finding $H /$ at site $i$. In the same equation, the second last term, also occurring with the same rate, corresponds to a transition from a local configuration $\backslash H /$ to $ / H \backslash$ and assuming the $H$ particle at site $i$, this process reduces the probability to find  $H /$ at site $i$. In this manner all the terms in the above set of equations can be interpreted. It can be easily verified that the sum of right hand side of all four equations is zero, as expected from the conservation of probability. We have not been able to solve these equations analytically and therefore solve them numerically for different values of $b,b',\rho$ and $m$ and use these solutions in the expression for $J_\rho$ and $J_m$.

\section{Master equation for three-point correlation functions}
\label{app:3pt}
In this appendix, we present the time-evolution equations for the three-point correlators like $P(/ H \backslash)$, $P(H / L)$, etc. There are $16$ possible variables like this and their time-evolution equations can be similarly constructed, following the steps outlined in the previous appendix. For example, the time evolution of $P(H \backslash L)$ can be written as
\begin{align}
\frac{dP(H \backslash L)}{dt} &=(1-\rho)(1-m)P(H \backslash H)+(1-\rho)(1-m)P(H \backslash L) \nonumber \\
&+(E-b^\prime)(1-m)P(L/H)+(E-b)(1-\rho)P(\backslash H/)\nonumber \\
&-P(H \backslash L)-(1-\rho)(1-m)P(L/H)-\rho m P(H \backslash L)\nonumber \\
&-(E+b^\prime)m P(H \backslash L)-(E+b)(1-\rho)P(/H \backslash). \label{eq:b1}
\end{align}
Here, the first term on the right hand side represents the process where a local configuration $(H \backslash H \backslash L)$ changes to $(H \backslash L \backslash H)$ and if the two bonds shown in this configuration are $i$-th and $(i+1)$-th, then such a transition increases the probability of finding an $HL$ pair across the $i$-th bond with a downward tilt. In this manner all the terms in the above equation can be explained. We write down the equations for the other $15$ quantities below. 
\begin{align}
\frac{dP(L \backslash H)}{dt} &=P(L \backslash H)+\rho(1-m)P(L/H)+\rho m P(L \backslash L) \nonumber \\
&+(E+b)(1-m)P(L/H)+(E+b^\prime)\rho P(\backslash L/) \nonumber \\
&-(1-\rho)(1-m)P(L \backslash H)-\rho(1-m) P(H \backslash L) \nonumber \\
&-(E-b) m P(L \backslash H)-(E-b^\prime)\rho P(/L \backslash) \\
~\nonumber \\
\frac{dP(H \backslash H)}{dt} &=\rho (1-m)P(H \backslash L)+\rho m P(H \backslash L)+(E+b)(1-m)P(H/H) \nonumber \\
&-(1-\rho)(1-m)P(H \backslash H)-\rho (1-m) P(L/H) \nonumber \\
&+(E-b)mP(H \backslash H)-(E+b)\rho P(/H \backslash) \\
~\nonumber \\
\frac{dP(L \backslash L)}{dt} &=(1-\rho)(1-m) P(L/H)+(1-\rho)(1-m)P(L \backslash H) \nonumber \\
&+(E-b^\prime)(1-m) P(L/L)+(E+b^\prime)(1-\rho)P(\backslash L/) \nonumber \\
&-(1-\rho)(1-m) P(H \backslash L)-\rho m P(L \backslash L)-(E+b^\prime)m P(L \backslash L) \nonumber \\
&-(E-b^\prime)(1-\rho)P(/L \backslash) 
\end{align}
\begin{align}
\frac{dP(L/H)}{dt} &=\rho m P(L/L)+\rho mP(L/H)+(E-b)m P(L \backslash H) \nonumber \\
&+(E-b^\prime)\rho P(/L \backslash)-P(L/H)-\rho)m P(H \backslash L) \nonumber \\
&-(1-\rho)(1-m) P(L/H)-(E+b^\prime)\rho P(\backslash L/)\nonumber \\
&-(E+b)(1-m)P(L/H) \\
~ \nonumber \\
\frac{dP(H/L)}{dt} &=P(L/H)+(1-\rho)(1-m)P(H/H)+(1-\rho)m P(H \backslash L) \nonumber \\
&+(E+b)(1-\rho) P(/H \backslash)+(E+b^\prime)m P(H \backslash L) \nonumber \\
&-(1-\rho)m P(L/H)-\rho m P(H/L)-(E-b^\prime)(1-m)P(H/L)\nonumber \\
&-(E+b)(1-\rho)P(\backslash H/) \\
~ \nonumber \\
\frac{dP(H/H)}{dt} &=\rho m P(H/L)+\rho mP(H \backslash L)+(E-b) m P(H \backslash H) \nonumber \\
&+(E+b)\rho P(/H \backslash)- \rho m P(L/H)-(1-\rho)(1-m) P(H/H) \nonumber \\
&-(E+b)(1-m) P(H/H)-(E-b)\rho P(\backslash H/) \\
~\nonumber \\
\frac{dP(L/L)}{dt} &=(1-\rho) m P(L/H)+(1-\rho)(1-m)P(L/H) \nonumber \\
&+(E-b^\prime)(1-\rho)P(/L \backslash)+(E+b^\prime)m P(L \backslash L)- \rho m P(L/L) \nonumber \\
&-(1-\rho)m P(H \backslash L)-(E-b^\prime)(1-m) P(L/L) \nonumber  \\
&-(E+b^\prime)(1-\rho)P(\backslash L/) 
\end{align}
\begin{align}
\frac{dP(/H \backslash)}{dt} &=(E-b)P(\backslash H/)+(E+b)\rho(1-m)P(/H \backslash) \nonumber \\
&+(E+b)(1-m)P(/H/)+(E-b^\prime)\rho(1-m)P(/L \backslash) \nonumber \\
&+(E-b^\prime)(1-\rho)(1-m)P(/H/)-(E+b)P(/H \backslash) \nonumber \\
&-(1-\rho)P(/H \backslash)-(1-m)P(L/H)-(E-b)\rho m P(/H \backslash) \nonumber \\
&-(E-b)\rho(1-m)P(\backslash H/)-(E+b^\prime)(1-\rho)m P(/H \backslash) \nonumber \\
&-(E+b^\prime\rho (1-m)P(\backslash L/) \\
~ \nonumber \\
\frac{dP(\backslash H/)}{dt} &=m P(H \backslash L)+\rho P(\backslash L/)+(E+b)P(/H \backslash) \nonumber \\
&+(E-b)\rho m P(\backslash H \backslash)+(E+b^\prime)\rho m P(\backslash L/) \nonumber \\
&-(E-b)P(\backslash H/)-(E+b)\rho(1-m)P(\backslash H/) \nonumber \\
&-(E+b)\rho(1-m)P(\backslash H/)-(E+b)\rho m P(/H \backslash) \nonumber \\
&-(E-b^\prime)(1-\rho)(1-m) P(\backslash H/)-(E-b^\prime)\rho m P(/L \backslash) \\
~\nonumber \\
\frac{dP(/L \backslash)}{dt} &=(1-\rho) P(/H \backslash)+(1-m)P(L/H)+(E+b^\prime)P(\backslash L/) \nonumber \\
&+(E-b^\prime)(1-\rho)(1-m) P(/L \backslash)+(E-b^\prime)(1-\rho)(1- m) P(/L/) \nonumber \\
&+(E+b)(1-\rho)(1-m)P(/H \backslash)+(E+b)\rho(1-m)P(/L/) \nonumber \\
&-(E+b^\prime)P(/L \backslash)-(E-b)\rho m P(/L \backslash) \nonumber \\
&-(E-b)(1-\rho)(1-m) P(\backslash H/)-(E+b^\prime)(1-\rho) m P(/L \backslash) \nonumber \\
&-(E+b^\prime)(1-\rho)(1-m)P(\backslash L/) \\
~\nonumber \\
\frac{dP(\backslash L/)}{dt} &=(E-b^\prime)P(/L \backslash)+(E-b)\rho mP(L \backslash L)+(E-b)(1-\rho) m P(\backslash H/) \nonumber \\
&+(E+b^\prime)(1-\rho)m P(\backslash L \backslash)+(E+b^\prime)(1-\rho)m P(\backslash L/) \nonumber \\
&-(E+b^\prime)P(\backslash L/)-\rho P(\backslash L/)-m P(H \backslash L) \nonumber \\
&-(E+b)\rho(1-m)P(\backslash L/)-(E+b)(1-\rho) m P(/H \backslash) \nonumber \\
&-(E-b^\prime)(1-\rho)(1-m) P(\backslash L/)-(E-b^\prime)(1-\rho) m P(/L \backslash)
\end{align}
\begin{align}
\frac{dP(/H/)}{dt} &=\rho P(/L/)+(E-b)\rho m P(/H \backslash)+(E+b^\prime)(1-\rho) m P(/H \backslash) \nonumber \\
&+(E+b)\rho m P(/H \backslash)+(E-b^\prime)\rho m P(/L \backslash)-mP(L/H) \nonumber \\
&-(E+b)\rho (1-m) P(/H/)-(E-b^\prime)(1-\rho)(1-m)P(/H/) \nonumber \\
&-(E-b)\rho m P(\backslash H/)-(E+b^\prime)\rho m P(\backslash L/)\\
~\nonumber \\
\frac{dP(/L/)}{dt} &=m P(L/H)+(E-b)\rho m P(/L \backslash)+(E+b^\prime)(1-\rho) m P(/L \backslash) \nonumber \\
&+(E+b)(1-\rho) m P(/H \backslash)+(E-b^\prime)(1-\rho) m P(/L \backslash)-\rho P(/L/) \nonumber \\
&-(E+b)\rho (1-m) P(/L/)-(E-b^\prime)(1-\rho)(1-m)P(/L/) \nonumber \\
&-(E-b)(1-\rho) m P(\backslash H/)-(E+b^\prime)(1-\rho) m P(\backslash L/)\\
~\nonumber \\
\frac{dP(\backslash H \backslash)}{dt} &=(1-m) P(H \backslash L)+(E-b)\rho (1-m) P(\backslash H/) \nonumber \\
&+(E+b^\prime)\rho(1-m) P(\backslash L/)+(E+b)\rho(1-m) P(\backslash H/) \nonumber \\
&+(E-b^\prime)(1-\rho) (1-m) P(\backslash H/)-(1-\rho) P(\backslash H \backslash) \nonumber \\
&-(E+b)\rho (1-m) P(/H \backslash)-(E-b^\prime)\rho(1-m)P(/L \backslash) \nonumber \\
&-(E-b)\rho m P(\backslash H \backslash)-(E+b^\prime)(1-\rho) m P(\backslash H \backslash)\\
~\nonumber \\
\frac{dP(\backslash L \backslash)}{dt} &=(1-\rho) P(\backslash H \backslash)+(E+b)\rho (1-m) P(\backslash L/) \nonumber \\
&+(E-b^\prime)(1-\rho)(1-m) P(\backslash L/) \nonumber \\
&+(E-b)(1-\rho)(1-m) P(\backslash H/)+(E+b^\prime)(1-\rho)(1-m) P(\backslash L/)\nonumber \\
&-(1-m) P(H \backslash L)-(E+b)(1-\rho)(1-m) P(/H \backslash) \nonumber \\
&-(E-b^\prime)(1-\rho)(1-m)P(/L \backslash)-(E-b)\rho m P(\backslash L \backslash) \nonumber \\
&-(E+b^\prime)(1-\rho) m P(\backslash L \backslash). 
\end{align}
In the steady state, when all time-derivatives are zero, these equations can be solved using Mathematica and a closed form expression for each three point correlator can be obtained. These expressions are too long to be presented here, but using them $J_\rho$ and $J_m$ can be calculated.


\section{ Comparison between numerical simulations and various approximate calculations for currents and wave speeds }
\label{app:tab}
\begin{table} [ht] 
\begin{center}
{\renewcommand\arraystretch{2}
\begin{tabular}{|c|c|c|c|c|c|c|c|c|c|}
\hline
$b,b^\prime$ & $\rho,m$ & \multicolumn{4}{c|}{$J_\rho$} & \multicolumn{4}{c|}{$J_m$} \\
\hline
&  & simulations & two point & three point & mean field & simulations & two point & three point & mean field \\
\hline
$-0.3,-0.3$ & $0.16,0.203$ & $0.073$ & $0.0734$ & $0.084$ & $0.0798$ & $0.0649$ & $0.0626$ & $0.0626$ & $0.066$ \\
\hline
$-0.4,-0.1$ & $0.12,0.26$ & $0.0424$ & $0.0421$ & $0.056$ & $0.0507$ & $0.0117$ & $0.0127$ & $0.0167$ & $0.0154$ \\
\hline
$-0.1,-0.1$ & $0.3,0.3$ & $0.0569$ & $0.0684$ & $0.089$ & $0.084$ & $0.0152$ & $0.0137$ & $0.0139$ & $0.0168$ \\
\hline
$-0.3,-0.3$ & $0.89,0.23$ & $0.048$ & $0.058$ & $0.061$ & $0.053$ & $-0.0824$ & $-0.085$ & $-0.078$ & $-0.083$ \\
\hline
$0.1,-0.3$ & $0.67,0.49$ & $0.0028$ & $0.019$ & $0.001$ & $0.0044$ & $0.09$ & $0.083$ & $0.084$ & $0.083$ \\
\hline
$0.1,-0.3$ & $0.85,0.34$ & $0.0254$ & $0.056$ & $0.034$ & $0.041$ & $0.063$ & $0.057$ & $0.062$ & $0.058$ \\
\hline
\end{tabular}}
\end{center}
\caption{Comparison between particle and tilt currents measured in simulations and calculated using approximations based on mean field theory, two point correlations and three point correlations.  The error bar in the numerically measured currents are $\pm 0.00005$.}
\label{tab1}
\end{table}

\begin{table} [ht] 
\begin{center}
{\renewcommand\arraystretch{2}
\begin{tabular}{|c|c|c|c|c|c|c|c|c|c|}
\hline
$b,b^\prime$ & $\rho,m$ & \multicolumn{4}{c|}{$\lambda_1$} & \multicolumn{4}{c|}{$\lambda_2$} \\
\hline
&  & simulations & two point & three point & mean field & simulations & two point & three point & mean field \\
\hline
$-0.5,0$ & $0.36,0.5$ & $-0.323$ & $-0.364$ & $-0.36$ & $-0.339$ & $0.323$ & $ 0.291$ & $0.305$ & $0.339$ \\
\hline
$-0.3,-0.5$ & $0.56,0.535$ & $-0.421$ & $-0.441$& $-0.454$ & $-0.442$ & $0.465$ & $0.441$ & $0.423$ & $0.444$ \\
\hline
$-0.3,-0.3$ & $0.89,0.23$ & $-0.563$ & $-0.545$ &$-0.541$& $-0.558$ & $-0.139$ & $-0.113 $ & $-0.109$ & $-0.116$ \\
\hline
$-0.3,-0.5$ & $0.895,0.2$ & $-0.612$ & $-0.608$ & $-0.615$ & $-0.611$ & $-0.136$ & $-0.118$ & $-0.112$ & $-0.122$ \\
\hline
$0.1,-0.3$ & $0.67,0.49$ & $-0.177$ & $-0.209$ & $-0.230$ & $-0.21$ & $0.178$ & $0.191$ & $0.176$ & $0.21$ \\
\hline
$0.1,-0.3$ & $0.85,0.34$ & $-0.226$ & $-0.271$ & $-0.288$ & $-0.286$ & $0.156$ & $0.142$ & $0.151$ & $0.145$ \\
\hline
\end{tabular}}
\end{center}
\caption{ Comparisons between the numerically measured speeds which we estimate from the moving peaks of the dynamical structure factors, and speeds calculated using the currents obtained using mean field theory or within other approximations where two point and three point correlation functions are retained. The error bars in the numerically measured values of wave speeds are less than $\pm 0.001$} \label{tab2}
\end{table}

\end{document}